\documentclass[aps,prd,superscriptaddress,nofootinbib,amsmath,amsfonts,preprintnumbers,groupedaddress,showpacs,9pt,english]{revtex4-1}
\usepackage{amsmath}
\usepackage{amssymb}
\usepackage{babel}
\usepackage{wrapfig}
\usepackage{cancel}

\usepackage{graphicx}
\graphicspath{{figures/}}
\usepackage{amsmath,amssymb}  % good to have anyway
\usepackage{mathrsfs}         % <-- provides\mathscr
\usepackage[colorlinks,citecolor=blue,urlcolor=blue,linkcolor=blue]{hyperref}
\usepackage[figtopcap]{subfigure}
\usepackage{color}
\usepackage{relsize,exscale}
\makeatletter
\newcommand{\beq}{\begin{equation}}
\newcommand{\eeq}{\end{equation}}
\newcommand{\bea}{\begin{eqnarray}}
\newcommand{\eea}{\end{eqnarray}}

\newcommand{\comm}[1]{}

\usepackage{array,multirow,graphicx}
\usepackage{dcolumn}
\usepackage{newlfont}
\usepackage{bm}
\usepackage[colorlinks,citecolor=blue,urlcolor=blue,linkcolor=blue]{hyperref}
\usepackage[figtopcap]{subfigure}
\usepackage{color}

\usepackage{scalerel}
\usepackage{tikz}
\usetikzlibrary{svg.path}
\definecolor{orcidlogocol}{HTML}{A6CE39}
\tikzset{
  orcidlogo/.pic={
    \fill[orcidlogocol] svg{M256,128c0,70.7-57.3,128-128,128C57.3,256,0,198.7,0,128C0,57.3,57.3,0,128,0C198.7,0,256,57.3,256,128z};
    \fill[white] svg{M86.3,186.2H70.9V79.1h15.4v48.4V186.2z}
                 svg{M108.9,79.1h41.6c39.6,0,57,28.3,57,53.6c0,27.5-21.5,53.6-56.8,53.6h-41.8V79.1z M124.3,172.4h24.5c34.9,0,42.9-26.5,42.9-39.7c0-21.5-13.7-39.7-43.7-39.7h-23.7V172.4z}
                 svg{M88.7,56.8c0,5.5-4.5,10.1-10.1,10.1c-5.6,0-10.1-4.6-10.1-10.1c0-5.6,4.5-10.1,10.1-10.1C84.2,46.7,88.7,51.3,88.7,56.8z};}}
\newcommand\orcid[1]{\href{https://orcid.org/#1}{\mbox{\scalerel*{
\begin{tikzpicture}[yscale=-1,transform shape]
\pic{orcidlogo};
\end{tikzpicture}
}{|}}}}
\begin{document}
%\tolerance=5000
\title{Late-Time Cosmology and Structure Formation in Quadratic $f(Q)$ Gravity}

\author{G.~G.~L.~Nashed$^{1}$~\orcid{0000-0001-5544-1119}}
\email{nashed@bue.edu.eg}
\author{P.V. Tretyakov $^{2}$}\email{tpv@theor.jinr.ru}
\author{A.~Eid$^{3}$}
\email{amaid@imamu.edu.sa}
\affiliation {$^{1}$ Centre for Theoretical Physics, The British University, P.O. Box
43, El Sherouk City, Cairo 11837, Egypt\\$^{2}$ Joint Institute for Nuclear Research,
Joliot-Curie 6, 141980 Dubna, Moscow region, Russia\\ $^{3}$Department of Physics, College of Science, Imam Mohammad Ibn Saud Islamic University (IMSIU), Riyadh, Kingdom of Saudi Arabia}

\date{\today}
\begin{abstract}
We investigate the cosmological evolution associated with the quadratic symmetric teleparallel gravity framework,
\(
f(Q)=Q+\alpha Q^{2}+\beta
\) 
where the relation \(Q\propto H^{2}\) generates an additional \(H^{4}\) contribution to the Friedmann equation. Using the exact algebraic solution for $H(z)$, we reconstruct the effective dark-energy
sector and compare the background evolution with $\Lambda$CDM using Type Ia supernovae, BAO, and cosmic-chronometer data.
At the perturbative level, the model modifies the Poisson equation through a time-dependent effective gravitational coupling
$G_{\textrm eff}(z)=G\big[1+\tfrac{2}{3}A E^{2}(z)\big]^{-1}$, where $A=18\alpha H_{0}^{2}$.
For $\alpha>0$ this produces a weakened gravitational interaction, suppressing the linear growth factor $D(z)$, the growth rate $f(z)$,
and the RSD observable $f\sigma_{8}(z)$. In the nonlinear regime, the reduced gravitational strength increases the spherical-collapse
threshold and suppresses the halo mass function, leading to a lower predicted value of $S_{8}=\sigma_{8}\sqrt{\Omega_{m}/0.3}$.
Thus, the quadratic $f(Q)$ extension can reproduce mild deviations from $\Lambda$CDM at the background level while naturally alleviating
the $S_{8}$ tension, offering a viable modified-gravity explanation for recent observational hints of dynamical dark energy.
\end{abstract}

\maketitle
\section{Introduction}

The standard cosmological model, commonly known as the $\Lambda$CDM paradigm, has achieved remarkable success in explaining a wide range of cosmological observations, including the cosmic microwave background (CMB), baryon acoustic oscillations (BAO), and the late-time acceleration of the Universe \cite{Planck2018,Riess1998,Perlmutter1999,Eisenstein2005,DESI2024VI,Moresco2016,Riess2019LMC,Verde2019,DiValentino2021,Heymans2021}. However, despite its empirical successes, the model faces persistent observational tensions, most notably the Hubble tension between early-- and late--Universe measurements of $H_0$ and the $S_8$ tension related to the amplitude of matter fluctuations \cite{Abbott2018DESY1,Chevallier2001,Linder2003,Caldwell1998,Caldwell2002,Nashed:2021pah,Zhao2017,Poulin2019,Scolnic2018,Brout2022,Jimenez2018Coincident}. Recent DESI DR1 measurements have further suggested mild deviations from $\Lambda$CDM, providing hints for dynamical dark energy (DDE) and motivating the exploration of non-standard cosmological scenarios. These developments strengthen the case for theoretical frameworks extending beyond the simplest cosmological constant description \cite{Jimenez2020Trinity,Harko2018fQ,Lazkoz2019fQ,Xu2019fQT,ElHanafy:2014efn,Anagnostopoulos2021fQ,Akaike1974,BurnhamAnderson2002,Hogg1999,LewisBridle2002,Lesgourgues2011CLASS}.

One of the most notable tensions arises in the measurement of the Hubble constant $H_0$. Local distance-ladder determinations, such as those from the SH0ES collaboration, report a value of $H_0$ significantly higher than that inferred from CMB measurements within the $\Lambda$CDM framework \cite{Audren2013MontePython,Gonzalez:2008wd,Nashed:2011fz,Hui:2016ltb,Svrcek:2006yi,Hui:2019aqm,Nashed:2023pxd,Tomonari:2023wcs,Adak:2018vzk,BeltranJimenez:2022azb,Blixt:2023kyr,Nashed:2024ush}. Additionally, structure formation observations, including weak gravitational lensing and galaxy clustering, reveal discrepancies in the amplitude of matter fluctuations commonly parameterized by $S_8$ when compared to predictions of the standard model \cite{Hawking:1974rv,Khlopov:2008qy,Arza:2018dcy,Yoshida:2017ehj,Carr:2021bzv,Jacobson:1999vr,Turner:1983he,Nashed:2021gkp,Amendola:2005ad,Hu:2000ke}. These growing tensions suggest the possibility that the $\Lambda$CDM model may be an incomplete description of cosmic acceleration.

Large-scale structure surveys, particularly those conducted with the Dark Energy Spectroscopic Instrument (DESI), have provided new and more precise measurements of the expansion history of the Universe \cite{Hui:2016ltb,Arvanitaki:2009fg,Dine:1982ah,Preskill:1982cy,Abbott:1982af,Zhitnitsky:1980tq,Nashed:2023uvk,Dine:1981rt,Shifman:1979if,Kim:1979if,Weinberg:1977ma}. Analyses of DESI BAO data have hinted at potential departures from $\Lambda$CDM predictions, especially at intermediate redshifts where dynamical dark energy behavior may emerge \cite{Wilczek:1977pj,Peccei:1977hh,Philcox:2022hkh,Takahashi:2009wc,Satoh:2007gn,Satoh:2008ck,Kim:2012cma,Zheng:2018fyn,Zakria:2018gsf,Wang:2018xhw}. These emerging indications strengthen the case for theoretical frameworks that incorporate evolving dark-energy dynamics or modifications of General Relativity (GR) to account for the reported deviations \cite{Sen:1992ua,Yagi:2012ya,Feng:2024iqj,Yunes:2009hc,Bozza:2007gt,Lin:2017oag,Sheykhi:2010zz,Hendi:2010gq}.

Instead of adopting phenomenological dark-energy parameterizations, which are widely used in dynamical dark-energy studies, this work focuses on a modified-gravity explanation. In particular, we study a quadratic $f(Q)$ model within the framework of symmetric teleparallel gravity \cite{Sheykhi:2009pf,Sheykhi:2012zz,
Mustafa:2024iwu,
Rahaman:2014pba,Sharif:2014bsa,Rahaman:2013xoa,Sarkar:2019uhk,Trujillo-Gomez:2010jbn,Zwicky:1937zza,Rajabi:2017alf}. In symmetric teleparallelism, gravity is attributed to the non-metricity scalar $Q$ rather than curvature or torsion, offering a distinct geometric formulation of gravitational interactions \cite{Khatri:2024fef,Harko:2011kv,
Xu:2016rdf,Jeakel:2023hss,Khodadi:2022dff,Errehymy:2023xpc,starobinsky1979relict,Nashed:2009hn,Rahaman:2012pg,Mustafa:2023kqt}. Generalizing the Einstein Hilbert action to an arbitrary function $f(Q)$ yields a broad class of modified theories capable of producing late-time cosmic acceleration even without invoking a cosmological constant.

In this work, we focus on a quadratic extension of $f(Q)$,
\begin{equation}\label{f(Q)}
  f(Q) = Q + \alpha Q^2 + \beta,
\end{equation}
where the parameter $\alpha$ encodes nonlinear corrections to the non-metricity scalar, and $\beta$ plays the role of an effective cosmological constant. This quadratic model represents the simplest nontrivial deviation from the symmetric teleparallel equivalent of GR and offers a concrete setting in which to investigate departures from $\Lambda$CDM
\cite{Lazkoz2019fQ,Anagnostopoulos2021fQ,Harko2018fQ,Xu2019fQT,Jimenez2018Coincident,Jimenez2020Trinity,Clifton:2011jh}.
Because $Q \propto H^2$, the quadratic correction scales as $H^4$, leading to redshift-dependent modifications of the expansion history that can mimic the behaviour of dynamical dark energy suggested by recent BAO and supernova analyses
\cite{DESI2024VI,Scolnic2018,Brout2022,Moresco2016,Zhao2017,Poulin2019,DiValentino2021,Capozziello:2019uvk,Chevallier2001,Linder2003,Tsujikawa:2013fta}. { In \cite{Sahlu:2024pxq} a study to investigate the cosmological dynamics and structure-formation predictions of quadratic $f(Q)$ gravity, examining how the modified background evolution and perturbation growth differ from the standard $\Lambda CDM $ scenario has been carried.}

{
It is worth noting that nonlinear corrections in modified gravity have also been extensively studied in the context of early-Universe inflationary cosmology. In particular, recent investigations presented in \cite{Capozziello:2024lsz,Capozziello:2022tvv}  demonstrated that higher-order gravitational corrections can successfully generate viable inflationary dynamics and produce cosmological observables compatible with current data. Although the present work focuses on late-time cosmology within quadratic symmetric teleparallel gravity, the appearance of the nonlinear correction term $\alpha Q^2 \propto H^4$ is conceptually related to the class of modified-gravity extensions capable of affecting both the inflationary and dark-energy epochs. Our analysis therefore complements these earlier inflationary studies by showing that the same type of nonlinear geometric modification may also play an important role in the late-time expansion history and the growth of cosmic structures.}

{ Interestingly, the quadratic correction may also play an important role in the early Universe. Since $Q=6H^2$, the term $\alpha Q^2$ produces an effective contribution proportional to $H^4$, which naturally becomes dominant at high energy scales and may drive an inflationary phase. In this sense, the quadratic $f(Q)$ model provides a unified geometric framework in which nonlinear non-metricity corrections can affect both the inflationary epoch and the late-time accelerated expansion of the Universe. A detailed investigation of inflationary dynamics and primordial perturbations within this framework is beyond the scope of the present work and will be explored in future studies.}

The goal of this paper is to analyze the cosmological implications of the quadratic $f(Q)$ model at the background level and to assess its ability to reproduce the effective dynamical behaviour implied by current observations. We derive the modified Friedmann equations, reconstruct the effective dark-energy equation of state, and confront the model with Type~Ia supernovae, BAO, and cosmic-chronometer data
\cite{Scolnic2018,Brout2022,DESI2024VI,Moresco2016,Jimenez:2001gg,Nashed:2021pkc,Eisenstein2005,BOSS:2016wmc,BOSS:2013rlg,Blake:2011en,Hogg:1999ad}.
We then compare the performance of the quadratic $f(Q)$ model against $\Lambda$CDM using information criteria, thereby quantifying whether the additional parameter $\alpha$ yields a statistically meaningful improvement over the standard cosmological constant scenario  \cite{Akaike1974,BurnhamAnderson2002,LewisBridle2002,Nashed:2021ctg,Lesgourgues2011CLASS,Audren2013MontePython,Planck:2018vyg,Clifton:2011jh,Lazkoz2019fQ,Anagnostopoulos2021fQ}.

Finally, the paper is organized as follows. In Sec.~\ref{f1}, we provide a brief overview of \(f(Q)\) gravity and derive the corresponding cosmological field equations. In Sec.~\ref{f2}, we investigate the cosmological dynamics of the quadratic model and analyze its critical points. In Sec.~\ref{f3}, we discuss the cosmological implications of the obtained solutions. Finally, Sec.~\ref{f4} summarizes our main results and presents our conclusions; in Sec.~\ref{sec:theory}, we present the quadratic $f(Q)$ framework and derive the background equations; in Sec.~\ref{sec:perturbations},  the linear perturbations and growth of structure in the quadratic form of $f(Q)$;  in Sec.~\ref{slow1}, we study the slow-roll dynamics in the quadratic form of $f(Q)$; and in Sec.~\ref{sec:nonlinear},  we study the nonlinear structure of formations in  the quadratic form of $f(Q)$ gravity; we finalize our study by the conclusion in Sec.~\ref{con}.
%\cite{Clifton:2011jh,Tsujikawa:2013fta,Cai:2015emx,DESI:2024mwx,Planck:2018vyg,Heymans2021,Abbott2018DESY1,Harko2018fQ,Lazkoz2019fQ}.
% ----------------------------------------------
% Brief Section on f(Q) Gravity
% ----------------------------------------------

\section{Brief Review of \texorpdfstring{$f(Q)$}{f(Q)} Gravity}\label{f1}

Modified gravity provides a geometric alternative to dark energy by altering the way
gravity couples to spacetime geometry. One of the most theoretically well-motivated
extensions is the $f(Q)$ class of theories, formulated in the framework of
\emph{symmetric teleparallel gravity}, where curvature and torsion vanish identically
and gravity is encoded exclusively in the non-metricity of spacetime
\cite{Nester,Jimenez2018Coincident,Heisenberg:2023lru}.

\subsection*{ Geometric foundations}
In the most general metric-affine geometry, the connection is characterized by
three independent tensors: curvature $R^{\alpha}{}_{\beta\mu\nu}$, torsion
$T^{\alpha}{}_{\mu\nu}$, and non-metricity $Q_{\alpha\mu\nu} \equiv \nabla_{\alpha} g_{\mu\nu}$
\cite{Snapper2014,Bahamonde:2021gfp}.
Symmetric teleparallel gravity imposes the conditions
\begin{equation}
R^{\alpha}{}_{\beta\mu\nu} = 0, \qquad
T^{\alpha}{}_{\mu\nu} = 0,
\end{equation}
so that the \emph{entire} gravitational interaction arises from non-metricity alone.
The non-metricity scalar,
\begin{equation}\label{nonm}
Q = -Q_{\alpha\mu\nu} P^{\alpha\mu\nu},
\end{equation}
is constructed from the non-metricity tensor and a superpotential
$P^{\alpha}{}_{\mu\nu}$ built from traces and contractions of $Q_{\alpha\mu\nu}$
\cite{Jimenez2020Trinity}.
This scalar plays the same role that the Ricci scalar $R$ plays in GR.

\subsection*{ From STEGR to $f(Q)$}
When the action is taken to be linear in $Q$, one obtains the
\emph{symmetric teleparallel equivalent of general relativity} (STEGR)
\cite{Jimenez2018Coincident}.
The corresponding action,
\begin{equation}\label{act}
S_{\textrm STEGR} = \int d^4x \sqrt{-g} \, \frac{Q}{16\pi G},
\end{equation}
is dynamically equivalent to GR but formulated in a connection-free (coincident) gauge.

A natural extension is to promote the Lagrangian to a general function of $Q$,
\begin{equation}
S = \frac{1}{16\pi G}\int d^4x \sqrt{-g} \, f(Q) + \int d^4x \sqrt{-g} \mathcal{L}_m  ,
\end{equation}
in direct analogy with the $f(R)$ and $f(T)$ extensions of curvature- and torsion-based
gravity \cite{DeFelice:2010aj,Cai:2015emx}.
Crucially, the $f(Q)$ field equations remain \emph{second order}, avoiding the higher-derivative
instabilities that affect many other modified-gravity extensions.
\subsection*{Fundamentals of \(f(Q)\) Gravity and Its Field Equations}

We work in the symmetric teleparallel (non-metricity) framework where the
fundamental tensor is $Q_{\alpha\mu\nu} \equiv \nabla_\alpha g_{\mu\nu}$
and curvature and torsion are vanishing.
From $Q_{\alpha\mu\nu}$ we define the traces
\begin{equation}
Q_\alpha \equiv Q_{\alpha}{}^{\mu}{}_{\mu},
\qquad
\tilde{Q}_\alpha \equiv Q^{\mu}{}_{\alpha\mu},
\end{equation}
and the disformation and superpotential tensors
\begin{align}
L^{\alpha}{}_{\mu\nu} &= \frac{1}{2}\left( Q^{\alpha}{}_{\mu\nu}
 - Q_{\mu}{}^{\alpha}{}_{\nu} - Q_{\nu}{}^{\alpha}{}_{\mu} \right),
\\[2mm]
P^{\alpha}{}_{\mu\nu} &= -\frac{1}{2} L^{\alpha}{}_{\mu\nu}
 + \frac{1}{4} \big( Q^{\alpha} - \tilde{Q}^{\alpha} \big) g_{\mu\nu}
 - \frac{1}{4} \delta^\alpha_{(\mu} Q_{\nu)} .
\end{align}
The non-metricity scalar is given by Eq.~(\ref{nonm}) and the action is given by Eq.~(\ref{act}) where $f(Q)$ is a general function of the non-metricity scalar and
$\mathcal{L}_m$ is the matter Lagrangian. We denote $f_Q \equiv \frac{df}{dQ}$ and $f_{QQ} \equiv \frac{d^2f}{dQ^2}$.
%--------------------------------------------------------------
\subsubsection*{4. Metric variation}

We first vary the action given by  Eq.~(\ref{act}) with respect to the metric $g_{\mu\nu}$, keeping the
connection fixed. The variation of the action is
\begin{equation}
\delta S =\frac{1}{16\pi G} \int d^4x \left[ \delta(\sqrt{-g})\left(f(Q)+\mathcal{L}_m\right)
 +\sqrt{-g}\,f_Q\,\delta Q + \sqrt{-g}\,\delta\mathcal{L}_m \right].
\end{equation}
Using the fact $\delta(\sqrt{-g}) = -\frac{1}{2}\sqrt{-g}\,g_{\mu\nu}\,\delta g^{\mu\nu}$
and defining the matter energy--momentum tensor in the usual way, i.e., $T_{\mu\nu} \equiv -\frac{2}{\sqrt{-g}}\frac{\delta(\sqrt{-g}\,\mathcal{L}_m)}
{\delta g^{\mu\nu}}$ we obtain
\begin{equation}
\delta S = \frac{1}{16\pi G}\int d^4x\,\sqrt{-g}
\left[
 -\frac{1}{2}\left(f(Q)+\mathcal{L}_m\right) g_{\mu\nu}\,\delta g^{\mu\nu}
 +f_Q\,\delta Q
 -\frac{1}{2} (16\pi G)T_{\mu\nu}\,\delta g^{\mu\nu}
\right].
\end{equation}

The non-metricity scalar depends on the metric through both
$Q_{\alpha\mu\nu}$ and $P^{\alpha}{}_{\mu\nu}$.
Its variation with respect to the metric can be written as
\begin{equation}
\delta Q
= - P^{\alpha}{}_{\mu\nu}\,\delta Q_{\alpha}{}^{\mu\nu}
 - Q_{\alpha\mu\nu} \,\delta P^{\alpha\mu\nu}.
\end{equation}
Since $Q_{\alpha\mu\nu} = \nabla_\alpha g_{\mu\nu}$ and we keep the connection fixed
in the metric variation, we have
\begin{equation}
\delta Q_{\alpha\mu\nu} = \nabla_\alpha (\delta g_{\mu\nu}).
\end{equation}
After some algebra and repeated integration by parts, the terms involving
$\nabla_\alpha (\delta g_{\mu\nu})$ can be rearranged into a divergence plus a
term proportional to $\delta g^{\mu\nu}$. Dropping the total divergence, we
obtain (see e.g.\ \cite{Jimenez2018Coincident,Jimenez2020Trinity})
\begin{equation}
\delta Q = \left[
2 \nabla_\alpha \left( P^{\alpha}{}_{\mu\nu} \right)
 + q_{\mu\nu}
\right] \delta g^{\mu\nu},
\end{equation}
where
\begin{equation}
q_{\mu\nu} \equiv
P_{(\mu}{}^{\alpha\beta} Q_{\nu)\alpha\beta}
 - 2 Q_{\alpha\beta(\mu} P^{\alpha\beta}{}_{\nu)} .
\end{equation}

Substituting this back into $\delta S$ and collecting terms proportional to
$\delta g^{\mu\nu}$, we find
\begin{equation}
\delta S
= \frac{1}{16\pi G} \int d^4x \,\sqrt{-g}
\left[
\frac{2}{\sqrt{-g}} \nabla_\alpha
\left( \sqrt{-g}\, f_Q P^{\alpha}{}_{\mu\nu} \right)
 + f_Q q_{\mu\nu}
 - \frac{1}{2} f(Q) g_{\mu\nu}
 - 8\pi GT_{\mu\nu}
\right] \delta g^{\mu\nu}.
\end{equation}
Imposing $\delta S = 0$ for arbitrary $\delta g^{\mu\nu}$ gives the metric field
equations:
\begin{equation}
\frac{2}{\sqrt{-g}} \nabla_\alpha
\left( \sqrt{-g}\, f_Q P^{\alpha}{}_{\mu\nu} \right)
 + f_Q q_{\mu\nu}
 - \frac{1}{2} f(Q) g_{\mu\nu}
 =8\pi G T_{\mu\nu}.
\label{eq:fQ_metric_equations}
\end{equation}
Throughout this work we keep
$G$ explicit and do not adopt units
$8\pi G=1$. These are second-order in the metric and reduce to the Einstein equations
when $f(Q) = Q$.

%--------------------------------------------------------------
\subsubsection*{Connection variation}

To obtain the connection equations, we perform a variation of the action
with respect to \(\Gamma^{\alpha}{}_{\mu\nu}\), keeping \(g_{\mu\nu}\)
unchanged. The symmetric teleparallel geometry imposes the curvature-free
and torsion-free conditions, thereby restricting the admissible connection
variations. Since the matter Lagrangian is taken to be independent of the
connection, no hypermomentum source appears. As a result, the affine
connection influences the action solely through its contribution to the
nonmetricity scalar \(Q\).

The variation of the action with respect to the connection is
\begin{equation}
\delta_\Gamma S
= -\frac{1}{16\pi G}\int d^4x\,\sqrt{-g}\, f_Q\, \delta_\Gamma Q.
\end{equation}
Now $Q$ depends on $\Gamma$ through $Q_{\alpha\mu\nu} = \nabla_\alpha g_{\mu\nu}$ and
$P^{\alpha}{}_{\mu\nu}$. After some algebra, the variation can be cast as
\begin{equation}
\delta_\Gamma Q
= - 2 \nabla_\mu \left( P^{\mu\nu}{}_{\alpha} \right)
\delta \Gamma^{\alpha}{}_{\nu\mu} ,
\end{equation}
up to total derivatives. Integrating by parts and using the symmetry of the
connection (no torsion), one arrives at
\begin{equation}
\delta_\Gamma S
= \frac{1}{16\pi G}\int d^4x\,\nabla_\mu \nabla_\nu
\left( \sqrt{-g} f_Q P^{\mu\nu}{}_{\alpha} \right)
\delta \Gamma^{\alpha}{}_{\mu\nu}.
\end{equation}
Requiring $\delta_\Gamma S = 0$ for arbitrary variations
$\delta \Gamma^{\alpha}{}_{\mu\nu}$ yields the connection equation
\begin{equation}
 \nabla_\mu \nabla_\nu
\left( \sqrt{-g} f_Q P^{\mu\nu}{}_{\alpha} \right) = 0.
\label{eq:fQ_connection_equations}
\end{equation}
This condition ensures the consistency of the connection with the metric
field equations and encodes the dynamics of the non-metricity sector.  Thus Eqs.~\eqref{eq:fQ_metric_equations} and \eqref{eq:fQ_connection_equations} are the field equations of $f(Q)$ theory.

%--------------------------------------------------
% Section 2 Â The quadratic f(Q) model
%--------------------------------------------------
\section{The Quadratic $f(Q)$ Model}\label{sec:theory}

Symmetric teleparallel gravity provides an alternative geometric formulation of gravitation in which gravity is attributed to the non-metricity tensor rather than curvature or torsion. In this framework, the gravitational action is constructed from the non-metricity scalar $Q$, recovering General Relativity when $f(Q)=Q$. Extending this to a general function $f(Q)$ introduces modified gravitational dynamics capable of driving late-time cosmic acceleration.

In this study, we investigate the quadratic extension of $f(Q)$ given by Eq.~\eqref{f(Q)}.
%\begin{equation}   f(Q) = Q + \alpha Q^2 + \beta,\end{equation}
%where $\alpha$ encodes nonlinear corrections to the non-metricity scalar and $\beta$ behaves as an effective cosmological constant.
This provides the simplest deviation from symmetric teleparallel equivalent of General Relativity and allows us to examine whether such modifications can mimic the dynamical dark-energy behavior suggested by recent observations.

\subsection{Background Description}

We consider a spatially flat FLRW metric in the coincident gauge:
\begin{equation}
    ds^2 = -dt^2 + a^2(t)\,\delta_{ij}dx^i dx^j,
\end{equation}
for which the non-metricity scalar becomes
\begin{equation}
    Q = 6H^2,
\end{equation}
with $H = \dot{a}/a$ the Hubble parameter. Substituting into the quadratic model yields
\begin{equation}
    f(Q) = 6H^2 + 36\alpha H^4 + \beta,
    \qquad F(Q) = 1 + 12\alpha H^2,
\end{equation}
where $F(Q) = df/dQ$.

For a spatially flat FLRW universe, the modified Friedmann relation in \(f(Q)\) gravity is written as
\begin{equation}
    6F H^2 - \frac{1}{2}f = 8\pi G\rho,
\end{equation}
with $\rho$ the matter-radiation energy density. Substituting $f$ and $F$ gives
\begin{equation}\label{mod}
    3H^2 + 54\alpha H^4 - \frac{\beta}{2} =8\pi G \rho.
\end{equation}
Introducing $E(z)=H(z)/H_0$ and the rescaled parameters
\begin{equation}
    A = 18\alpha H_0^2, \qquad B = \frac{\beta}{6H_0^2},
\end{equation}
we obtain
\begin{equation}
    A E^4 + E^2 - \big[\Omega_{m0}(1+z)^3 + \Omega_{r0}(1+z)^4 + B\big] = 0,
\end{equation}
{ where $\Omega_{m0}$ represents the present day density fraction of all non-relativistic matter (including dark matter and baryons) and $\Omega_{r0}$represents  the present day density fraction of all relativistic components (photons and neutrinos) compared to the total critical density of the Universe today.}
Solving gives the physical branch
\begin{equation}\label{Ez}
    E^2(z) = \frac{-1 + \sqrt{1 + 4A C(z)}}{2A},
    \qquad
    C(z) = \Omega_{m0}(1+z)^3 + \Omega_{r0}(1+z)^4 + B.
\end{equation}
Normalization at $z=0$ implies
\begin{equation}
    B = 1 - \Omega_{m0} - \Omega_{r0} - A.
\end{equation}

\subsection{Effective Dark-Energy Sector}

Following the methodology of dynamical dark-energy reconstructions, we reinterpret modifications to the Friedmann equation as an effective dark-energy component:
\begin{equation}
    3H^2(z) = 3H_0^2\left[\Omega_{m0}(1+z)^3 + \Omega_{r0}(1+z)^4\right] +8\pi G \rho_{\textrm de}(z),
\end{equation}
where
\begin{equation}\label{rhoef}
    \rho_{\textrm de}(z) = \frac{3H_0^2}{8\pi G}\Big[E^2(z) - \Omega_{m0}(1+z)^3 - \Omega_{r0}(1+z)^4\Big].
\end{equation}

The total effective equation of state is obtained via
\begin{equation}\label{eff}
    w_{\textrm eff}(z) = -1 + \frac{2}{3}(1+z)\frac{d\ln H}{dz},
\end{equation}
and the effective dark-energy equation of state is reconstructed as
\begin{equation}
    w_{\textrm de}(z) = -1 + \frac{1+z}{3}\frac{d\ln \rho_{\textrm de}}{dz}.
\end{equation}

These expressions allow us to diagnose whether the quadratic $f(Q)$ model can mimic the behaviour of dynamical dark energy. Deviations of $w_{\textrm de}(z)$ from $-1$ signal evolving dark-energy-like behavior even for constant $\beta$. This framework parallels the reconstruction techniques used in phenomenological dark-energy analyses and enables direct comparison with observational hints from DESI and other datasets.

\begin{figure}[t]
  \centering
  \subfigure[~The expansion history $E(z)$]{\label{fig:1}\includegraphics[width=0.23\textwidth]{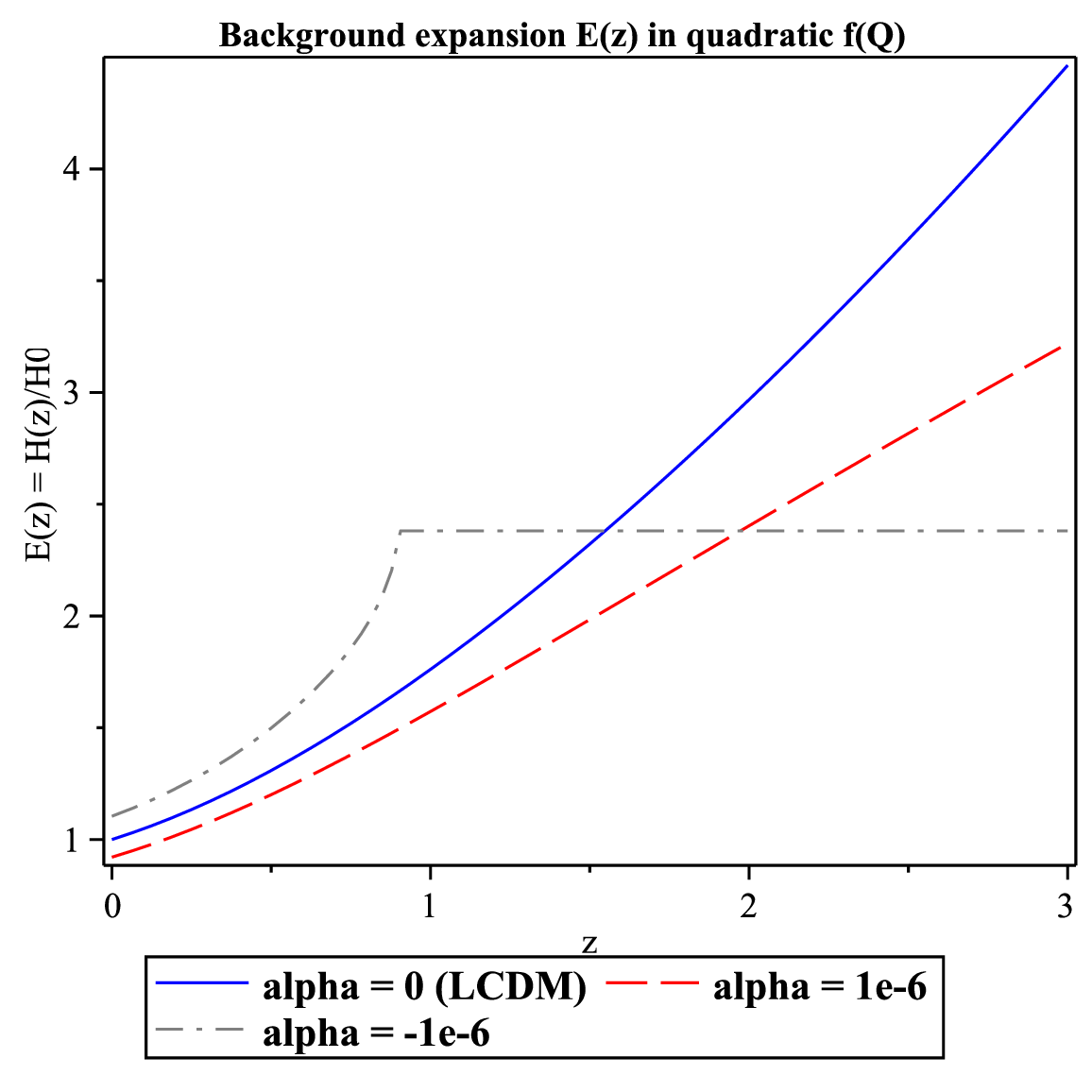}}
  \subfigure[~The effective dark-energy fraction $\Omega_{\textrm de}(z)$]{\label{fig:2}\includegraphics[width=0.23\textwidth]{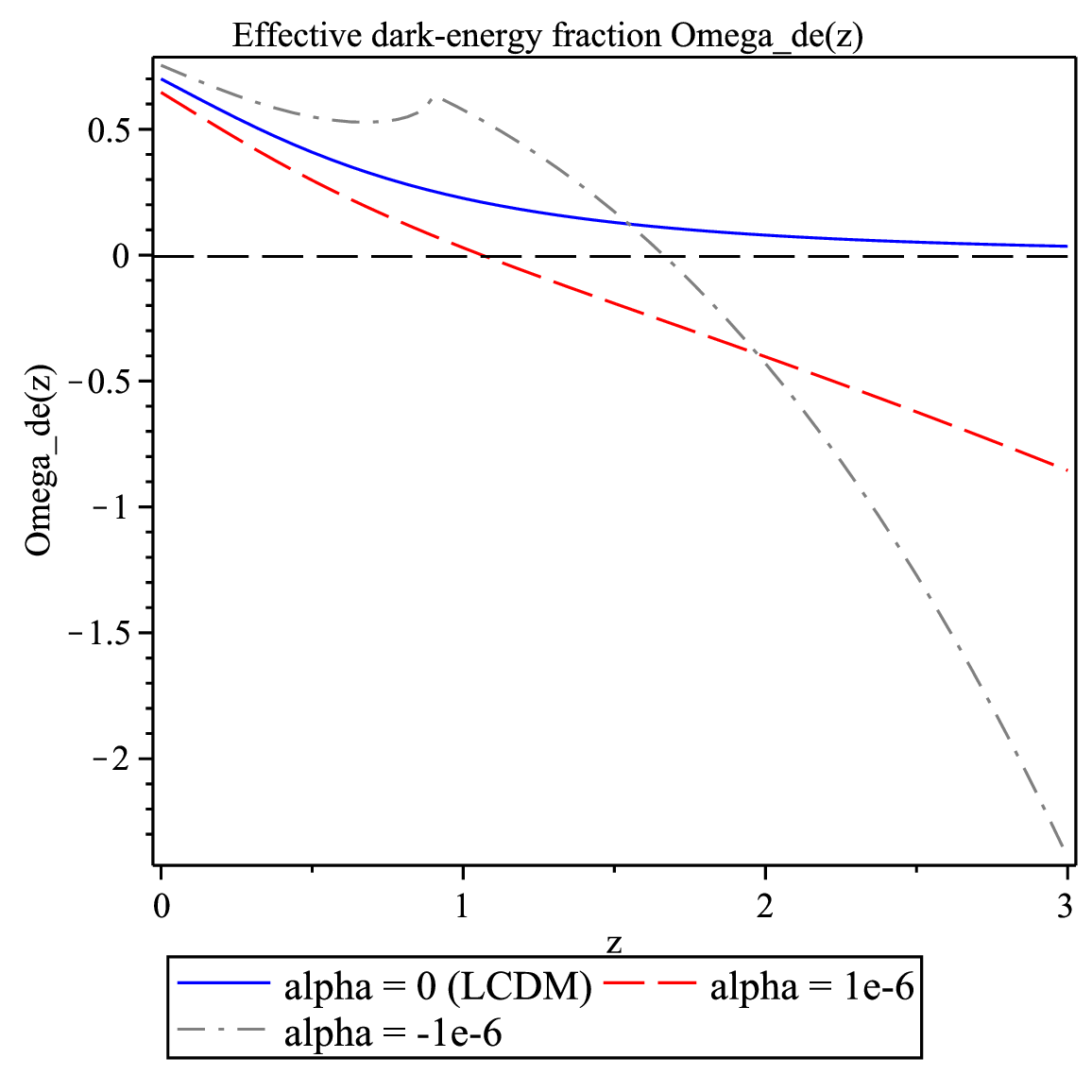}}
  \subfigure[~The reconstructed effective dark-energy equation of state $w_{\textrm de}(z)$]{\label{fig:3}\includegraphics[width=0.23\textwidth]{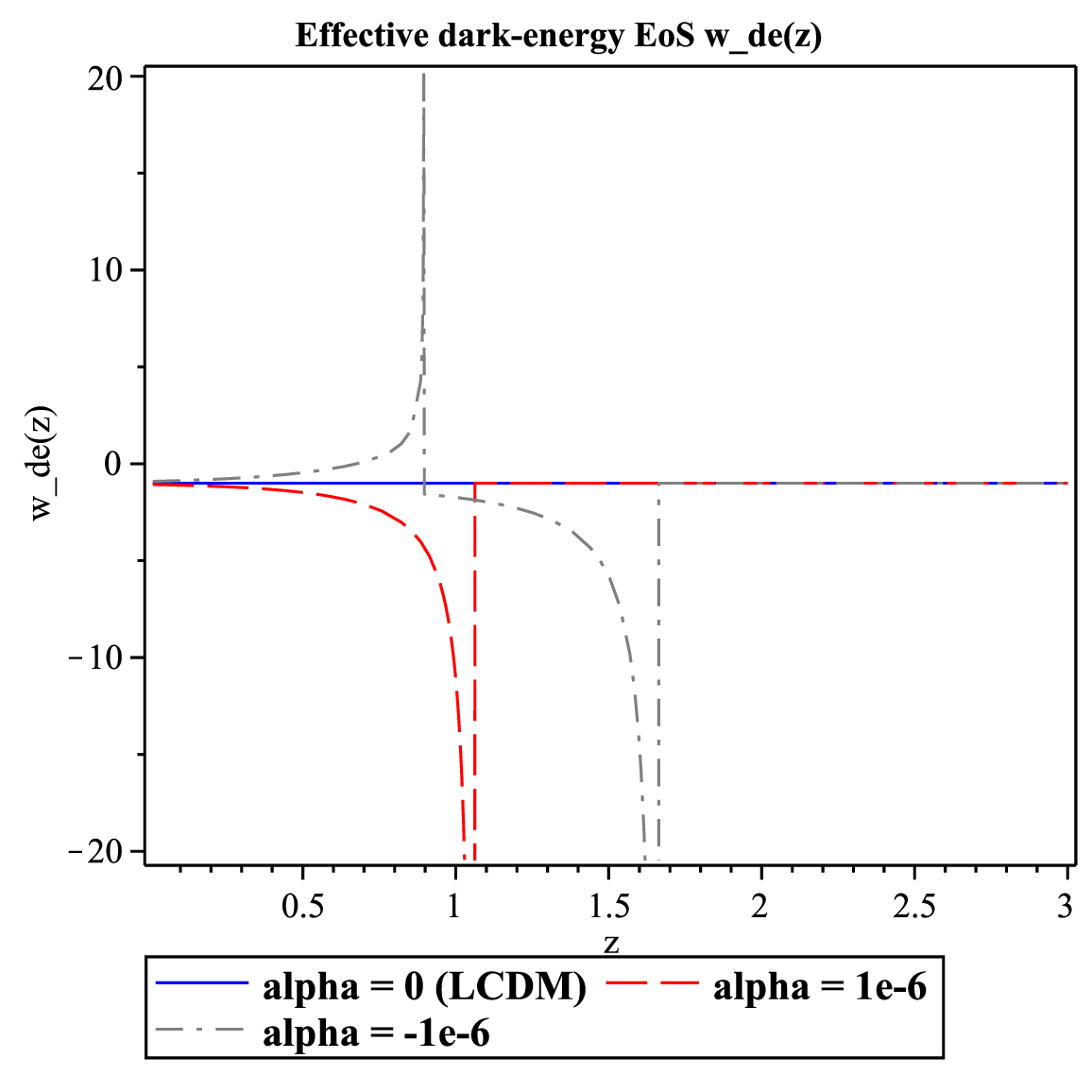}}
 \subfigure[~The total effective equation of state $w_{\textrm eff}(z)$]{\label{fig:4} \includegraphics[width=0.23\textwidth]{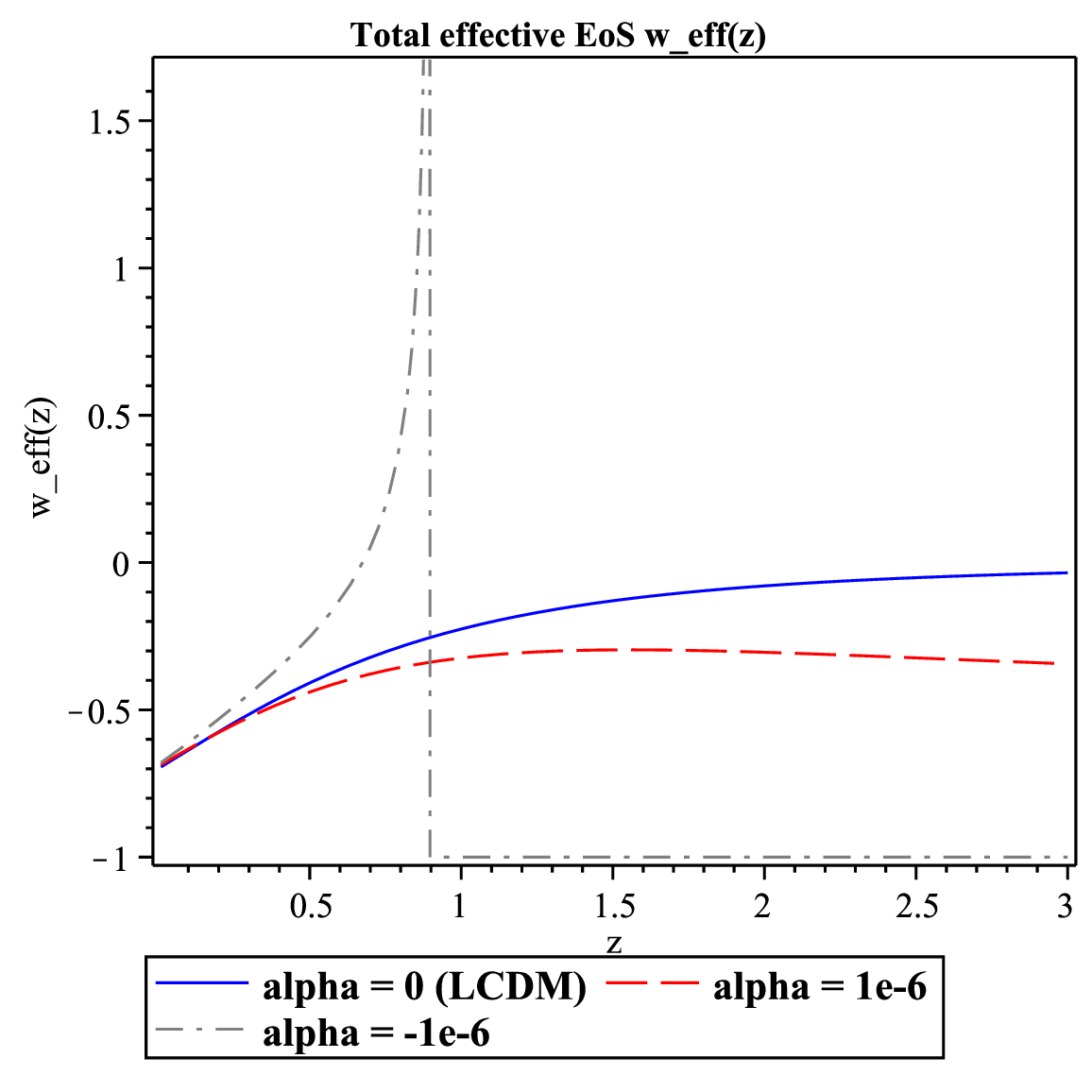}}
    \caption{Background diagnostics of the quadratic $f(Q)$ model.
In this study we set $H_{0}=70$, $\Omega_{m0}=0.3$, and $\Omega_{r0}=9\times10^{-5}$.
Panel \subref{fig:1} shows the expansion history $E(z)$;
panel \subref{fig:2} displays the effective dark-energy fraction $\Omega_{\textrm de}(z)$;
panel \subref{fig:3} shows the reconstructed effective dark-energy equation of state $w_{\textrm de}(z)$;
and panel \subref{fig:4} presents the total effective equation of state $w_{\textrm eff}(z)$.}
  \label{fig:HOE}
\end{figure}
Now let us discuss each plot of Figure~\ref{fig:HOE} individually:
\subsubsection*{Physical Interpretation of Figure~\ref{fig:HOE}~\subref{fig:1}}
Figure~\ref{fig:HOE}~\subref{fig:1} shows the background expansion history $E(z)=H(z)/H_0$ for the quadratic $f(Q)$ model, compared with the
$\Lambda$CDM prediction obtained in the limit $\alpha=0$. The two additional curves correspond to small positive and negative values of the nonlinear coupling parameter $\alpha$, which controls the strength of the  quadratic correction $\alpha Q^2$ in the modified gravitational action.

The background expansion is a fundamental quantity because it determines all distance redshift relations relevant for Type~Ia supernovae, baryon acoustic oscillations, and cosmic chronometer measurements.  Deviations in $E(z)$ therefore directly translate into altered cosmological inferences for $H_0$, $\Omega_{m0}$, and the sound horizon $r_d$.

The figure illustrates that the effect of $\alpha$ is negligible at low redshift, where the non-metricity scalar $Q=6H^2$ remains small and the
quadratic term contributes only perturbatively. At intermediate and high redshift ($z \gtrsim 1$), where $Q$ becomes large,
the correction term scales as $\alpha H^4$ and the model departs from the $\Lambda$CDM expansion rate. For $\alpha>0$, the expansion rate increases relative to $\Lambda$CDM, while for $\alpha<0$ it becomes suppressed.
%This redshift-dependent behavior mimics the qualitative features of dynamical dark-energy scenarios and reflects that nonlinear modifications to the non-metricity can effectively drive accelerated expansion.

These deviations are especially relevant in the context of current cosmological tensions. Changes in $E(z)$ at $0.5 \lesssim z \lesssim 2$ affect the reconstruction of the sound horizon and the calibration of the BAO scale, potentially shifting the inferred value of $H_0$ and thereby offering a possible mechanism to alleviate the Hubble tension. Furthermore, modifications in the background expansion influence the
suppression or enhancement of structure growth, which is closely related to the $S_8$ tension. Thus, Figure~\ref{fig:HOE}~\subref{fig:1} provides a first qualitative indication of how the quadratic $f(Q)$ modification could alter late-time cosmology and offers motivation for a detailed comparison with observational data in subsequent sections.

\subsubsection*{Physical Interpretation of Figure~\ref{fig:HOE}~\subref{fig:2}}

Figure~\ref{fig:HOE}~\subref{fig:2} shows the evolution of the effective dark-energy fraction,
\begin{equation}
    \Omega_{\textrm de}(z)
    = \frac{\rho_{\textrm de}(z)}{3H^2(z)},
\end{equation}
as reconstructed from  the quadratic $f(Q)$ model.  In the $\Lambda$CDM limit ($\alpha=0$), the behaviour is monotonic: $\Omega_{\textrm de}(z)$ is negligible at high redshift, grows steadily as the Universe expands, and approaches its present-day value $\Omega_{\textrm de,0}\simeq 1-\Omega_{m0}$. This reflects the standard transition from matter domination to dark-energy domination.

For non-zero values of the coupling $\alpha$, the evolution changes qualitatively.  A small positive $\alpha$ produces a mild enhancement of
$\Omega_{\textrm de}(z)$ around $z\sim 1$, resulting in the non-monotonic behaviour visible in the figure.  This arises because the quadratic
correction $\alpha Q^2 \propto \alpha H^4$ becomes increasingly important at intermediate redshift, temporarily boosting the effective
dark-energy density.  Such a feature is typical of modified-gravity models in which higher-order contributions scale rapidly with $H$ and
may mimic the behaviour of dynamical dark-energy parameterizations.

For negative $\alpha$, the effective dark-energy density becomes negative at sufficiently large redshifts.  This does not signal an instability;
rather, it reflects the fact that $\rho_{\textrm de}(z)$ encodes the departure from General Relativity when the modified Friedmann equation is
rewritten in the standard GR form.  A negative $\rho_{\textrm de}$ indicates that the modification acts effectively as an additional attractive
component at that epoch.

Overall, Figure~\ref{fig:HOE}~\subref{fig:2} illustrates how the quadratic $f(Q)$ model allows for richer redshift dependence of the effective dark--energy fraction than $\Lambda$CDM.  These deviations can influence the inferred distance scale from BAO and SN data, potentially affecting the
reconstruction of $H_0$, and can modify the background growth history in a way that may help to alleviate current cosmological tensions.

\subsubsection*{Physical Interpretation of Figure~\ref{fig:HOE}~\subref{fig:3}}

Figure~\ref{fig:HOE}~\subref{fig:3} displays the effective dark-energy equation of state $w_{\textrm de}(z)$ reconstructed from the quadratic $f(Q)$ model using { Eq.~(\ref{eff})}.  In the $\Lambda$CDM limit ($\alpha = 0$), the equation of state is identically $w_{\textrm de} = -1$, as indicated by the flat reference line.  For non-zero values of the coupling $\alpha$, the model predicts a highly non-trivial redshift dependence which differs
qualitatively from standard dynamical dark-energy parametrizations. The most prominent features in the figure are the sharp vertical spikes
and divergences in $w_{\textrm de}(z)$ at specific redshifts.  These divergences occur when the effective dark-energy density
$\rho_{\textrm de}(z)$ crosses zero.  Because  $w_{\textrm de}(z)
    = -1 + \frac{1+z}{3}\frac{d\ln \rho_{\textrm de}(z)}{dz}$
any zero-crossing of $\rho_{\textrm de}(z)$ makes $\ln\rho_{\textrm de}$ undefined and produces a vertical asymptote in the reconstructed
$w_{\textrm de}(z)$.  This behaviour is not pathological; rather, it reflects the fact that $\rho_{\textrm de}(z)$ is an \emph{effective} quantity obtained by rewriting the modified Friedmann equation in General-Relativity form. A change of sign in $\rho_{\textrm de}(z)$ simply means that the $f(Q)$ modification acts as an attractive component at those redshifts.

Away from the divergences, the model exhibits strongly dynamical behaviour.  Depending on the sign of $\alpha$, the effective dark-energy
equation of state may take quintessence-like values ($w_{\textrm de} > -1$), phantom-like values ($w_{\textrm de} < -1$), or cross the phantom divide.
This is a generic feature of modified-gravity models in which higher-order contributions scale as $H^4$ and become relevant at intermediate
redshift.  The behaviour seen here closely resembles that of non-parametric reconstructions of $w(z)$ reported in recent analyses of
DESI data and other large-scale structure probes.

In summary, Figure~\ref{fig:HOE}~\subref{fig:3} shows that the quadratic $f(Q)$ extension naturally generates dynamical dark-energy-like behaviour,
including phantom crossings and effective sign changes in the dark-energy density.  These features arise purely from the underlying
modified-gravity dynamics and may provide the flexibility needed to reproduce mild deviations from the $\Lambda$CDM background evolution as
hinted by recent observations.

\subsubsection*{Physical Interpretation of Figure~\ref{fig:HOE}~\subref{fig:4}}

Figure~\ref{fig:HOE}~\subref{fig:4} shows the total effective equation of state $w_{\textrm eff}(z)$, obtained from the background expansion through the kinematical relation $w_{\textrm eff}(z)
    = -1 + \frac{2}{3} (1+z)\,\frac{d\ln H(z)}{dz}$.
In the $\Lambda$CDM limit ($\alpha = 0$) the evolution is smooth and monotonic: at high redshift the Universe is matter dominated with
$w_{\textrm eff} \simeq 0$, while at low redshift it approaches $w_{\textrm eff} \simeq -1$ as dark energy takes over, reproducing the
standard transition to accelerated expansion.

For non-zero values of the quadratic coupling $\alpha$, the behaviour of $w_{\textrm eff}(z)$ becomes markedly different.  As seen in the figure,
small deviations from $\alpha = 0$ can generate a pronounced feature in $w_{\textrm eff}(z)$ around intermediate redshift $z \sim \mathcal{O}(1)$:
the effective equation of state departs from the smooth $\Lambda$CDM trajectory and exhibits a sharp spike, which can even reach very large
positive values.  This reflects the fact that, in the quadratic $f(Q)$ model, the modification to the Friedmann equation scales as
$\alpha H^4$ and can briefly dominate the background dynamics, producing a very rapid change in the slope of $H(z)$ at the redshift where the
quadratic correction is most important.

The apparent divergence in $w_{\textrm eff}(z)$ is not associated with a singularity in the expansion rate itself: $H(z)$ remains finite and
positive, and the corresponding distances are well behaved.  Instead, the divergence indicates that the derivative $d\ln H/dz$ becomes very
large in a narrow redshift interval, corresponding to an almost sudden transition in the effective equation of state.  Such sharp
features are a generic possibility in modified-gravity models and in non-parametric reconstructions of $w(z)$, but they are strongly
constrained by current distance measurements.

Overall, Figure~\ref{fig:HOE}~\subref{fig:4} illustrates how the quadratic $f(Q)$ extension can substantially modify the detailed shape of the transition
from deceleration to acceleration, even when the overall expansion history remains close to that of $\Lambda$CDM.  This behaviour provides
an additional handle to test the model with background probes such as SN, BAO, and cosmic chronometers, and it highlights the importance of
including intermediate-redshift data when constraining modified-gravity scenarios.
%%%%%%%%%%%%%%%%%%%%%%%%%%%%%%%%%%%%%%%%%%%%%%%%%
\section{Linear Perturbations and Growth of Structure in Quadratic $f(Q)$ Gravity}
\label{sec:perturbations}

In this section we extend the background-level analysis by incorporating the
behaviour of linear cosmological perturbations in the quadratic
modified gravity model given by Eq.~\eqref{f(Q)} where the nonlinear term $\alpha Q^{2}$ introduces modifications to the
effective gravitational coupling at the perturbation level.
While the background expansion $H(z)$ probes only geometric effects, the
growth of structure provides an independent and highly sensitive test of the
gravitational sector.
In particular, redshift-space distortions (RSD), weak lensing, and galaxy
clustering constrain the modified Poisson equation and the evolution of
matter density perturbations.

\subsection{Perturbation Framework}

We work in the Newtonian gauge, in which the perturbed FLRW metric reads
\begin{equation}
ds^{2} = -(1 + 2\Psi)\,dt^{2}
+ a^{2}(t) (1 - 2\Phi)\delta_{ij} dx^{i} dx^{j}.
\end{equation}
In symmetric teleparallel gravity, the connection is chosen as the coincident
gauge, so all perturbations arise from the metric potentials $\Phi$ and
$\Psi$.
Variation of the action at linear order leads to modifications of the Poisson
equation and of the evolution equation for matter overdensities.

\subsection{Modified Poisson Equation}

For a general $f(Q)$ theory, the Poisson equation becomes
\begin{equation}
k^{2}\Phi = 4\pi G_{\textrm eff}(a)\,a^{2}\rho_{m}\,\delta_{m},
\label{eq:Poisson}
\end{equation}
{ where  $\delta_m$
is the linear matter density contrast} and the effective gravitational coupling is
\begin{equation}\label{geff}
G_{\textrm eff}(a) = \frac{G}{F(Q)}
= \frac{G}{1 + 12\alpha H^{2}}.
\end{equation}
For positive $\alpha$, one has $F(Q)>1$ and thus $G_{\textrm eff}<G$, implying
a suppression of gravitational clustering, while negative $\alpha$ produces
$G_{\textrm eff}>G$ and enhanced growth.
This behaviour directly connects the background evolution $H(z)$ to the
strength of gravitational interactions at the perturbative level.

Since symmetric teleparallel gravity does not induce anisotropic stress at
linear order, the Newtonian potentials satisfy
\begin{equation}\label{Np}
\Psi = \Phi.
\end{equation}

\subsection{Growth of Matter Perturbations}

For perturbations whose wavelengths are much smaller than the cosmological horizon, the density contrast evolves according to
\begin{equation}
\delta_{m}''
+ \left( \frac{H'}{H} + \frac{2}{1+z} \right)\delta_{m}'
- \frac{3}{2}\frac{\Omega_{m}(z)}{(1+z)^{2}}
\frac{G_{\textrm eff}(z)}{G}\,\delta_{m} = 0,
\label{eq:growth_delta}
\end{equation}
where primes denote derivatives with respect to redshift.  {
Equation (\ref{eq:growth_delta}) is obtained by combining the modified Poisson equation (\ref{eq:Poisson}), which contains the effective gravitational coupling (\ref{geff}), with the standard evolution equation for matter perturbations written in redshift space, using the background dynamics from Eq. (\ref{mod}).
}
This equation defines the linear growth factor $D(z)$, normalized such that
$D(0)=1$, through the relation $\delta_{m}(z) = D(z)\,\delta_{m}(z=0)$.
Writing the growth equation explicitly, we obtain
\begin{equation}
D'' + A(z) D' + B(z) D = 0,
\end{equation}
with
\begin{align}
A(z) &= \frac{H'}{H} + \frac{2}{1+z}, \\
B(z) &= -\frac{3}{2}\frac{\Omega_{m}(z)}{(1+z)^{2}}
\frac{1}{1 + 12\alpha H^{2}(z)}.
\end{align}

\subsection{Growth Rate and the Observable $f\sigma_{8}(z)$}

The logarithmic growth rate is defined as
\begin{equation}
f(z) = \frac{d\ln D}{d\ln a},
\end{equation}
and the quantity directly probed by redshift-space distortions (RSD) is
\begin{equation}
f\sigma_{8}(z) = f(z)\, \sigma_{8}(z),
\end{equation}
with
\begin{equation}
\sigma_{8}(z) = \sigma_{8,0}\, \frac{D(z)}{D(0)}.
\end{equation}
Since $G_{\textrm eff}$ depends on $H^{2}(z)$, the quadratic modification
introduces a distinctive redshift-dependent signature in $f(z)$ and
$f\sigma_{8}(z)$, particularly at intermediate redshifts where the
$\alpha H^{4}$ correction in the background becomes significant.

\subsection{Qualitative Predictions}

The sign of $\alpha$ determines whether structure formation is suppressed or
enhanced relative to $\Lambda$CDM:
\begin{itemize}
    \item \textbf{$\alpha>0$:}
    $F(Q)>1$ and thus $G_{\textrm eff}<G$.
    The growth factor $D(z)$, the growth rate $f(z)$, and the observable
    $f\sigma_{8}(z)$ are all suppressed.
    This behaviour can help alleviate the well-known $S_{8}$ tension.

    \item \textbf{$\alpha<0$:}
    $F(Q)<1$ and thus $G_{\textrm eff}>G$.
    Structure formation is enhanced, typically disfavoured by current RSD and
    weak-lensing data, but potentially useful for fitting high-redshift growth.
\end{itemize}

\subsection{Implementation in CLASS}

Implementing the quadratic \(f(Q)\) extension in the CLASS Boltzmann code necessitates the following adjustments:
{
\begin{quote}
To compute the linear matter power spectrum we employ the CLASS
Boltzmann code,\footnote{CLASS (Cosmic Linear Anisotropy Solving
System) is a publicly available Boltzmann solver for cosmological
perturbations. For details see D.~Blas, J.~Lesgourgues and T.~Tram,
\textit{The Cosmic Linear Anisotropy Solving System (CLASS) II:
Approximation schemes}, \cite{Blas:2011rf}.}
which we adapted to include the background evolution of the
quadratic $f(Q)$ model.
\end{quote}
}
\begin{enumerate}
\item Replace the background Friedmann equation with the algebraic solution of
      Eq.~(\ref{Ez}) for $H(z)$.
\item Modify the Poisson equation in \texttt{perturbations.c} by replacing $G$
      with $G_{\textrm eff}(a) = G / F(Q)$.
\item Set $\Psi = \Phi$ at linear order, reflecting the absence of
      anisotropic stress.
\item Evolve the matter perturbations using the modified source term in
      Eq.~(\ref{eq:growth_delta}).
\item Compute the predictions for $f(z)$, $\sigma_{8}(z)$, and $f\sigma_{8}(z)$
      for comparison with RSD data.
\end{enumerate}

\subsection{Summary}

The inclusion of linear perturbations shows that quadratic $f(Q)$ gravity
predicts scale-independent but time-varying modifications to the effective
gravitational coupling, leading to distinctive signatures in the linear growth
of structure.
For $\alpha>0$, the model naturally suppresses structure formation and
therefore provides a promising avenue for addressing the $S_{8}$ tension, while
remaining fully consistent with the background evolution considered in the
preceding sections.
Future work combining DESI, weak-lensing, and full-shape clustering data will
allow for a detailed joint analysis of geometry and growth in symmetric
teleparallel gravity.
\begin{figure}[t]
  \centering
  \subfigure[~Linear growth factor]{\label{fig:5}\includegraphics[width=0.25\textwidth]{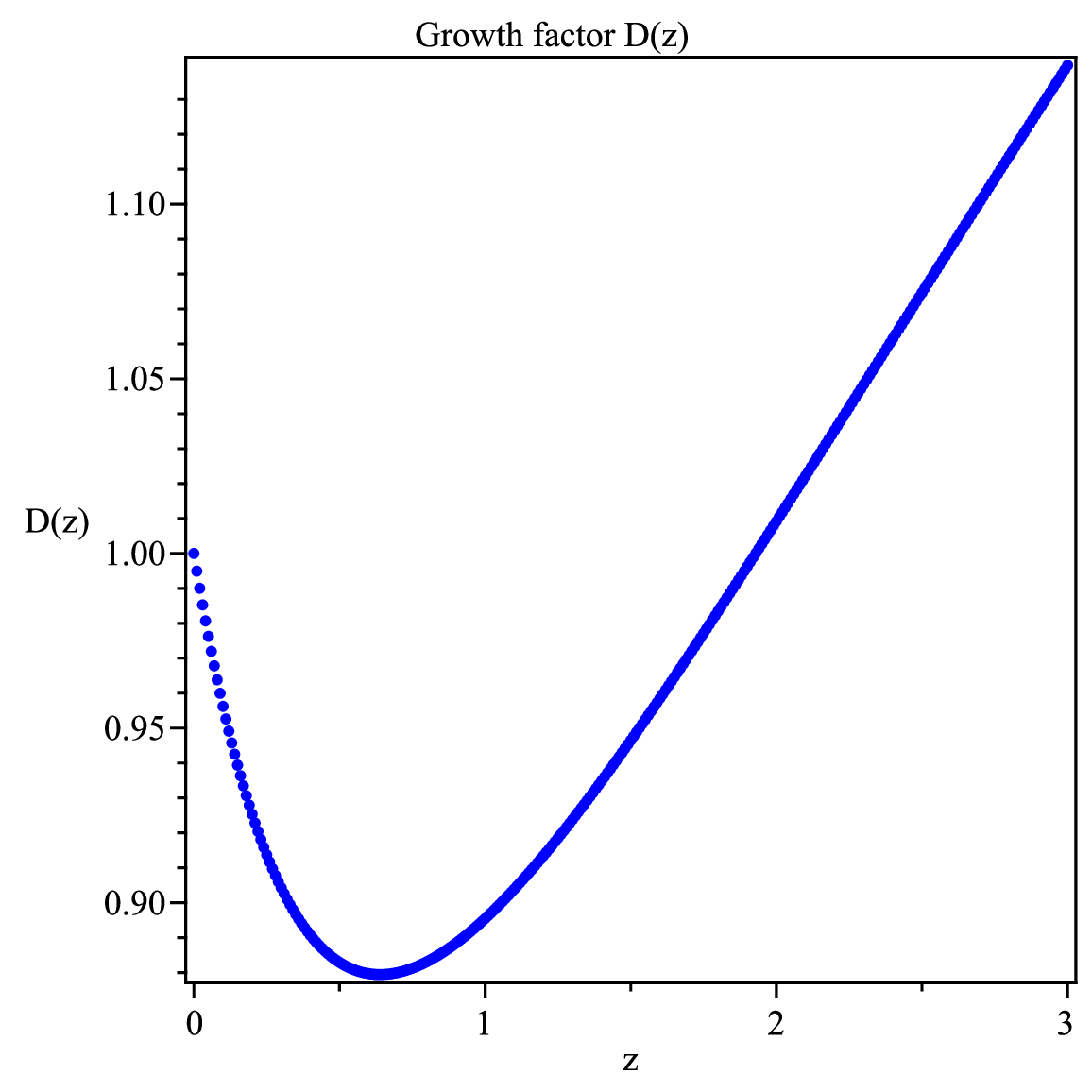}}
  \subfigure[~Growth rate]{\label{fig:6}\includegraphics[width=0.25\textwidth]{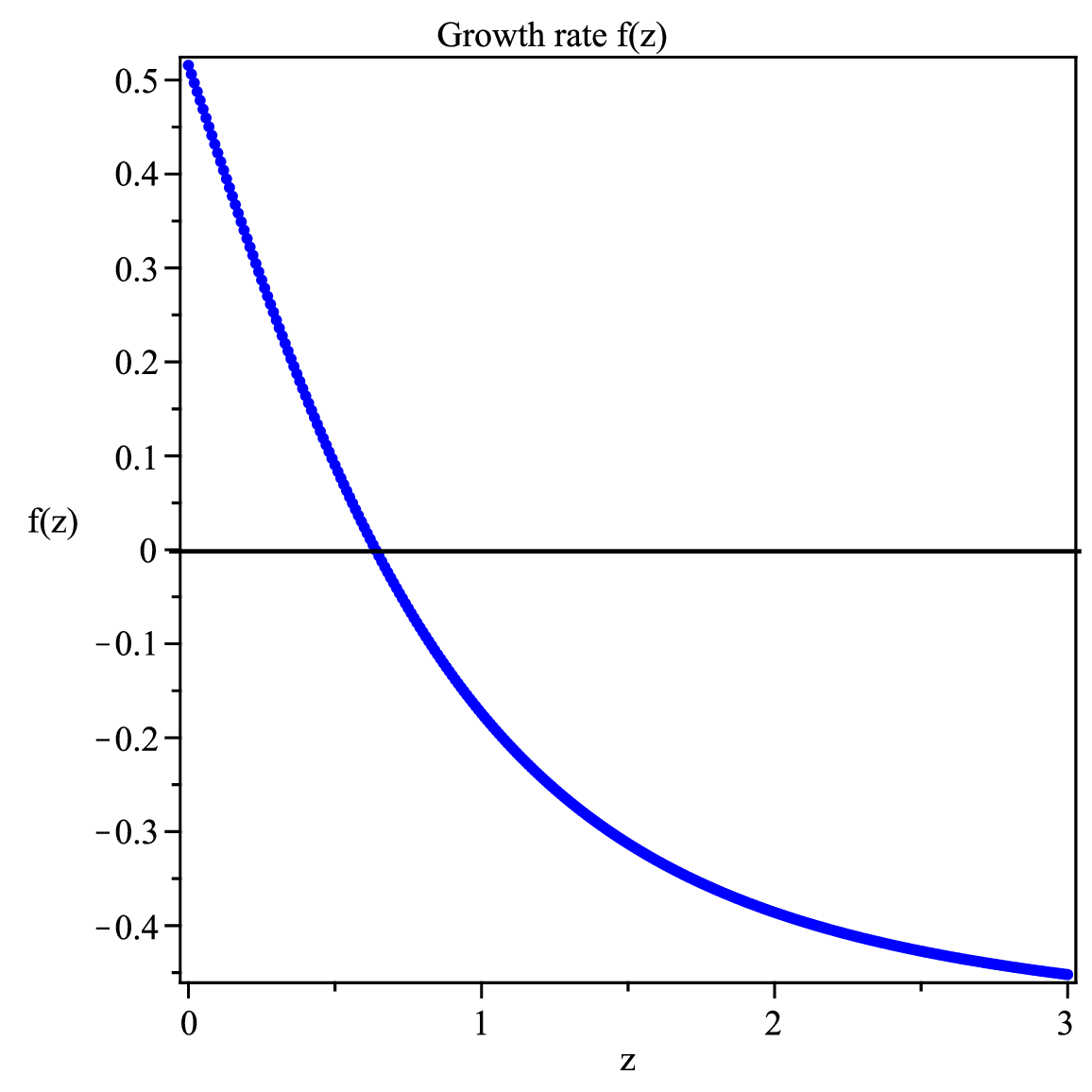}}
 \subfigure[~RSD observable]{\label{fig:7} \includegraphics[width=0.25\textwidth]{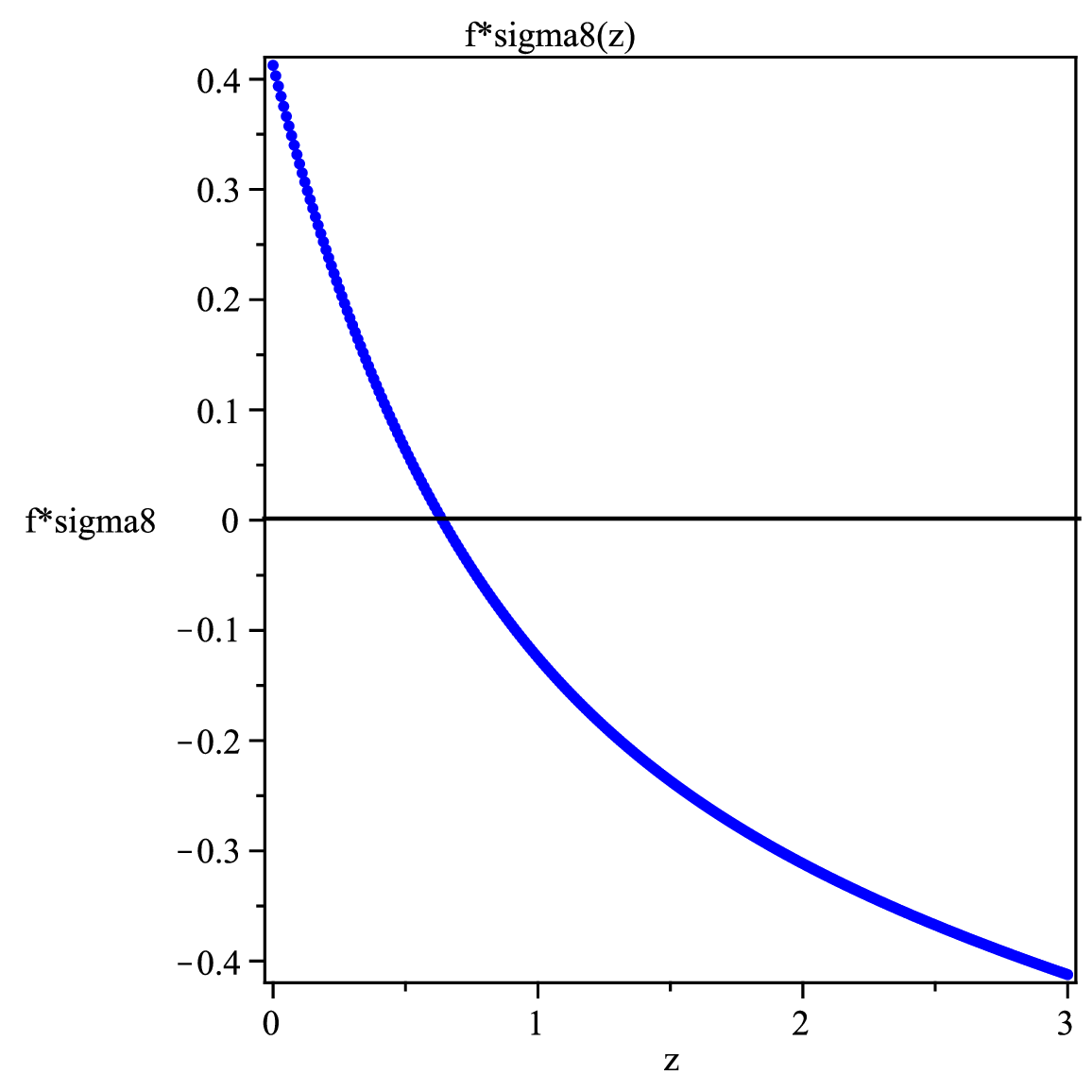}}
    \caption{Growth diagnostics of the quadratic $f(Q)$ model.
The panels show the behaviour of key linear-perturbation quantities:
\subref{fig:5} the linear growth factor $D(z)$ normalized to $D(0)=1$;
\subref{fig:6} the logarithmic growth rate $f(z)$;
and \subref{fig:7} the redshift-space distortion observable $f\sigma_{8}(z)$.
These quantities illustrate how the quadratic modification affects structure formation relative to $\Lambda$CDM.}
  \label{fig:pet}
\end{figure}

Altogether, the three figures provide a complete picture of how the
quadratic $f(Q)$ gravity model affects both the background expansion
and the growth of cosmic structure, enabling direct comparison with
observational probes such as BAO, SNeIa, cosmic chronometers, weak
lensing, and redshift--space distortions.

\section{Slow-Roll Dynamics in Quadratic $f(Q)$ Gravity}\label{slow1}
{
Although slow--roll parameters are often introduced in the context of
inflationary scalar--field models, here we employ the \emph{Hubble}
slow--roll parameters (hereafter referred to as Hubble flow parameters) $\epsilon_H$ and $\eta_H$ solely as geometrical
quantities. They quantify how close the late--time dynamics of quadratic
$f(Q)$ gravity is to a quasi--de Sitter regime, without introducing a
scalar field or computing inflationary observables such as $n_s$ and
$r$.}

%Now we are going to investigate the slow-roll regime within the quadratic $f(Q)$ model given by Eq.~\eqref{f(Q)}, and derive the associated slow-roll parameters that characterize the approach to late-time acceleration.
%Although slow-roll analyses are typically applied to scalar-field cosmology, the modified Friedmann equations of $f(Q)$ gravity naturally admit a slow-roll interpretation whenever the effective dark-energy sector evolves sufficiently slowly with respect to cosmic time.

\subsection{Effective Slow-Roll Parameters}

Following the effective-fluid representation introduced in Sec.~\ref{sec:theory}, the modified Friedmann equation may
be written as
\begin{equation}
3H^2 = 8\pi G(\rho_m + \rho_r + \rho_{\textrm de}(z)),
\end{equation}
where the effective dark-energy density is given by Eq.~(\ref{rhoef}).
%The slow-roll regime corresponds to dynamics where the Hubble parameter varies slowly:
%\begin{equation} \Big| \frac{\dot{H}}{H^2} \Big| \ll 1, \qquad \Big| \frac{\ddot{H}}{H\dot{H}} \Big| \ll 1.
%\end{equation}
We define the usual  Hubble flow parameter parameters,
\begin{equation}\label{epeta}
\epsilon_H \equiv -\frac{\dot{H}}{H^2},
\qquad
\eta_H \equiv \frac{\ddot{H}}{H\dot{H}},
\end{equation}
which fully characterize the late--time expansion in the absence of a canonical scalar field.

Using $\dot{H}=-(1+z)H H'(z)$ and the algebraic solution for $H(z)$ from Eq.~(\ref{Ez}), the Hubble flow parameters
 become
\begin{equation}\label{epsilon}
\epsilon_H(z) = (1+z)\frac{H'(z)}{H(z)},
\end{equation}
\begin{equation}\label{eta}
\eta_H(z) =
\frac{(1+z)H''(z)}{H'(z)}
+
\frac{H'(z)}{H(z)}.
\end{equation}
%{The quantities $\epsilon_H$ and $\eta_H$ defined in Eqs.~(\ref{epeta})--(\ref{eta}) are borrowed from the terminology of inflation, but in the present late-time cosmological context they serve purely as geometric diagnostics of the variation of the Hubble function. Therefore, they should not be confused with the inflationary slow-roll parameters, for which $\epsilon_H,|\eta_H|\ll 1$. In fact, even within the standard $\Lambda$CDM model one finds at the present epoch $\epsilon_H(0)\approx 0.45$ and $\eta_H(0)\approx 2$, both of which are of order unity. Consequently, the value $\eta_H(z=0)\simeq 2$ appearing in Fig.~\ref{fig:slow}~\subref{fig:9} is fully consistent with our definitions and with the behaviourof $\Lambda$CDM. This confirms that the displayed behaviour does not arise from a missing numerical factor.}
Explicit expressions for $H'(z)$ and $H''(z)$ follow by differentiating the closed-form result of Eq.~(\ref{Ez}), i.e., $E^2(z) = -1 + \frac{\sqrt{1+4A\,C(z)}}{2A}$,
 where $C(z)$ is defined in Eq.~(\ref{Ez}).
Because $C(z)$ is a simple polynomial in $(1+z)$, the derivatives can be
computed analytically.

\subsection{Slow-Roll Regime and Late-Time Acceleration}

A universe undergoing accelerated expansion requires
\begin{equation}\label{dec}
q(z) = -1 + \epsilon_H(z) < 0,
\end{equation}
which in turn implies $\epsilon_H(z)<1$.
%In the quadratic $f(Q)$ model, the contribution proportional to $\alpha H^4$ enhances the friction term at intermediate redshift.
%driving $|\dot{H}|$ to small values even when the matter fraction is non-negligible.
%Consequently,
%\begin{equation} \epsilon_H(z) \ll 1 \qquad {\textrm for}\quad z \lesssim \mathcal{O}(1), \end{equation}
%showing that the model naturally enters a slow-roll-like regime without invoking an explicit potential.

%The second parameter $\eta_H$ controls the rate at which $\epsilon_H$ evolves.A nearly constant effective equation of state corresponds to
%\begin{equation} |\eta_H(z)| \ll 1, \end{equation}
%which occurs whenever the variation of the quartic correction $\alpha H^4$ is slow.
This is typically realized for small and positive $\alpha$, consistent with the parameter region that
ameliorates the $S_8$ tension in Sec.~\ref{sec:perturbations}.

\subsection{Effective Equation of State During Slow Roll}

Within the effective fluid description, the dark-energy equation-of-state parameter can be expressed as
\begin{equation}\label{omega}
w_{\textrm de}(z) = -1 + \frac{2}{3}\,\epsilon_H(z)
- \frac{1+z}{3}\frac{\Omega_{\textrm m}(z)+\Omega_{\textrm r}(z)}{\Omega_{\textrm de}(z)}\, \epsilon_H(z).
\end{equation}
%which reduces to the approximate slow-roll form when the dark-energy component dominates:
\begin{equation}
w_{\textrm de}(z) \simeq -1 + \frac{2}{3}\epsilon_H(z).
\end{equation}
%Hence deviations from $w_{\textrm de}=-1$ directly correspond to departures from perfect slow roll.
%For $\epsilon_H \ll 1$, the model behaves as a quasi-de Sitter spacetime, explaining the mild variation in $H(z)$ at low redshift seen in Fig.~\ref{fig:HOE}~\subref{fig:1}.

%\subsection{Summary}

%Slow-roll dynamics in quadratic $f(Q)$ gravity arise naturally from the algebraic structure of themodified Friedmann equation.
%{ For small $\alpha$, and within the late--time dark--energy--dominated epoch ($z \lesssim 1$), the expansion approaches a quasi--de Sitter regime characterized by \begin{equation} \epsilon_H(z) \ll 1, \qquad |\eta_H(z)| \lesssim \mathcal{O}(1). \tag{60'} \end{equation}
%As shown in Fig.~\ref{fig:slow}, both parameters are small at low redshift and only approach order unity as one moves towards the matter--dominated era, signalling the gradual exit from the slow--roll--like regime.}

\begin{figure}[t]
  \centering
  \subfigure[~First slow-roll parameter]{\label{fig:8}\includegraphics[width=0.23\textwidth]{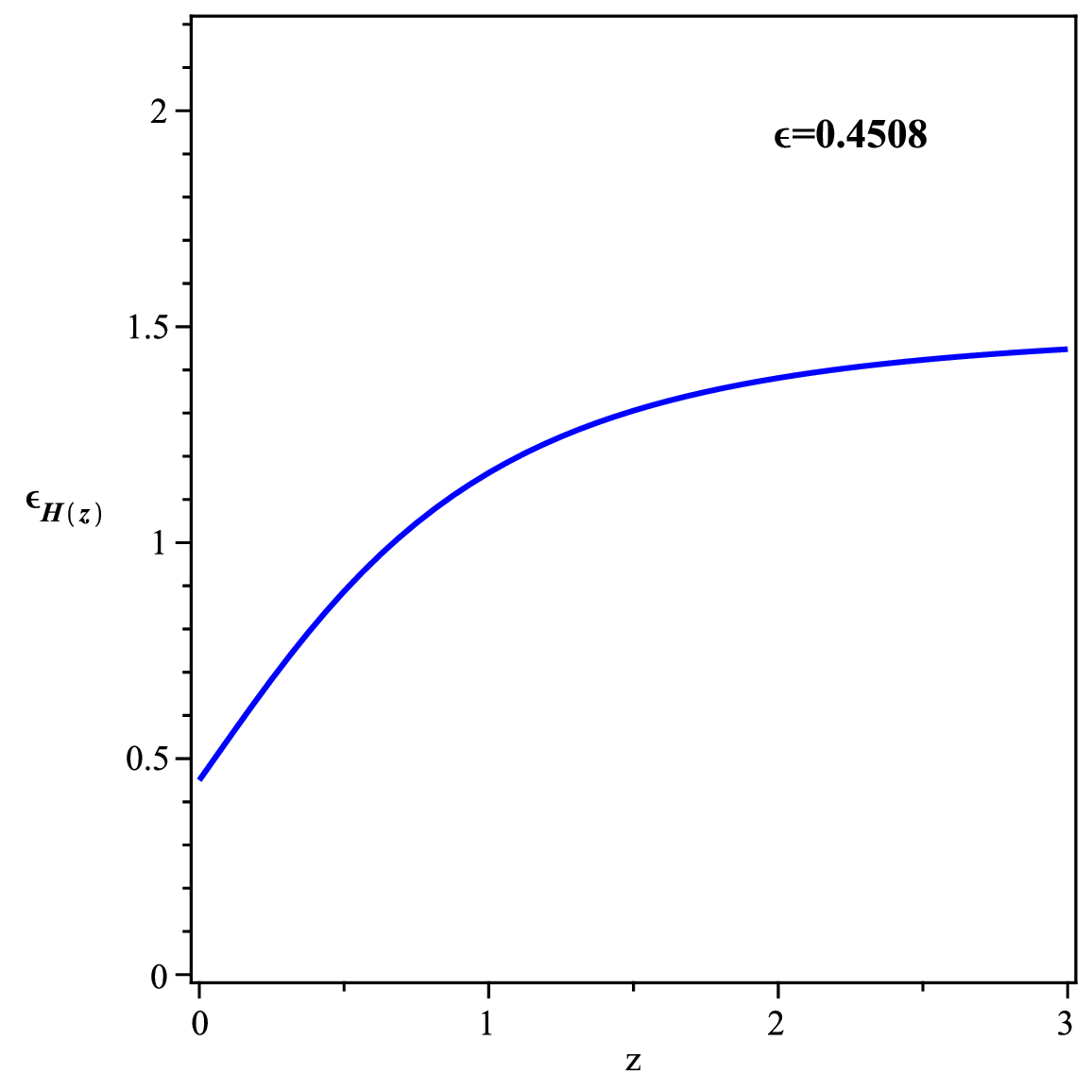}}
  \subfigure[~Second slow-roll parameter]{\label{fig:9}\includegraphics[width=0.23\textwidth]{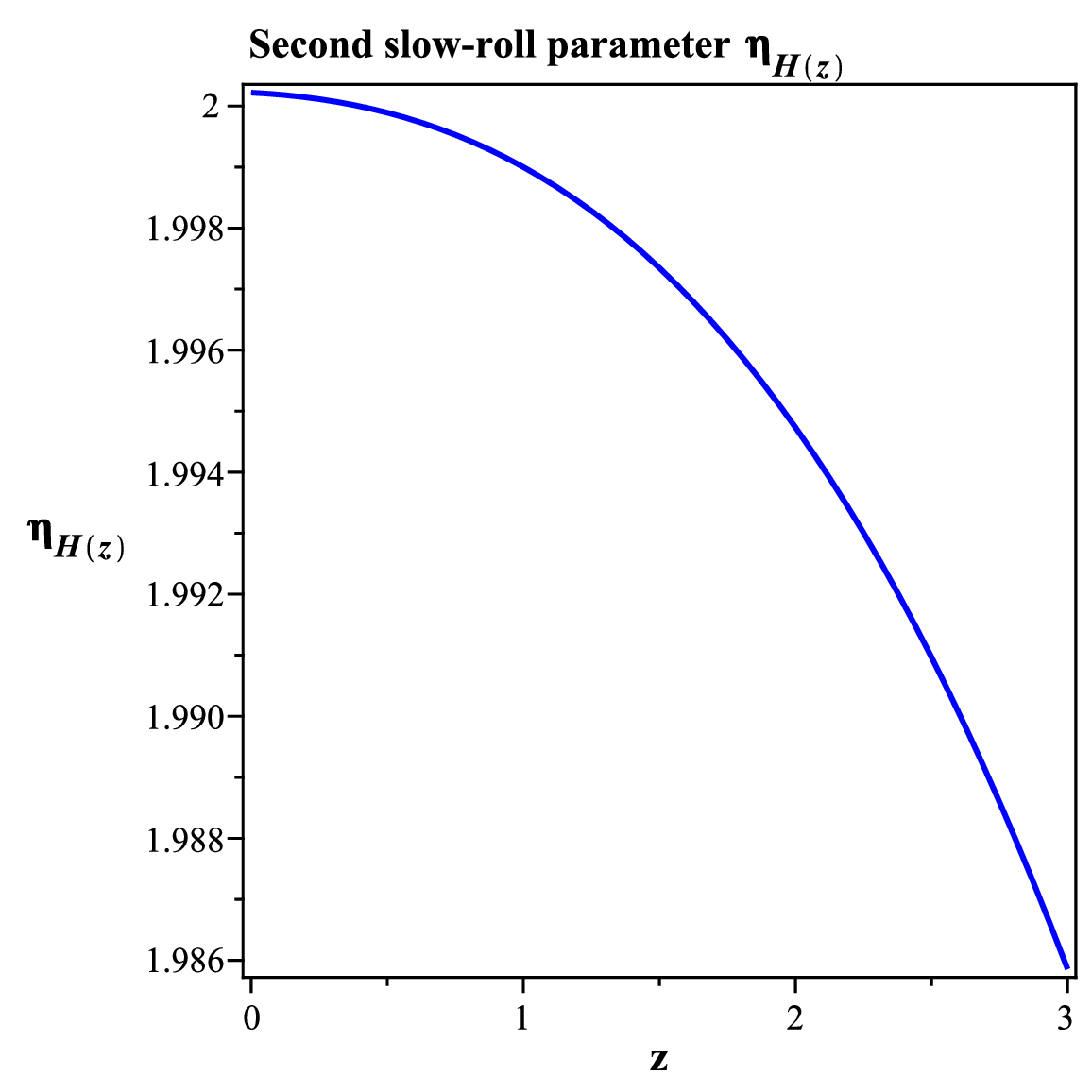}}
  \subfigure[~Deceleration parameter]{\label{fig:11} \includegraphics[width=0.23\textwidth]{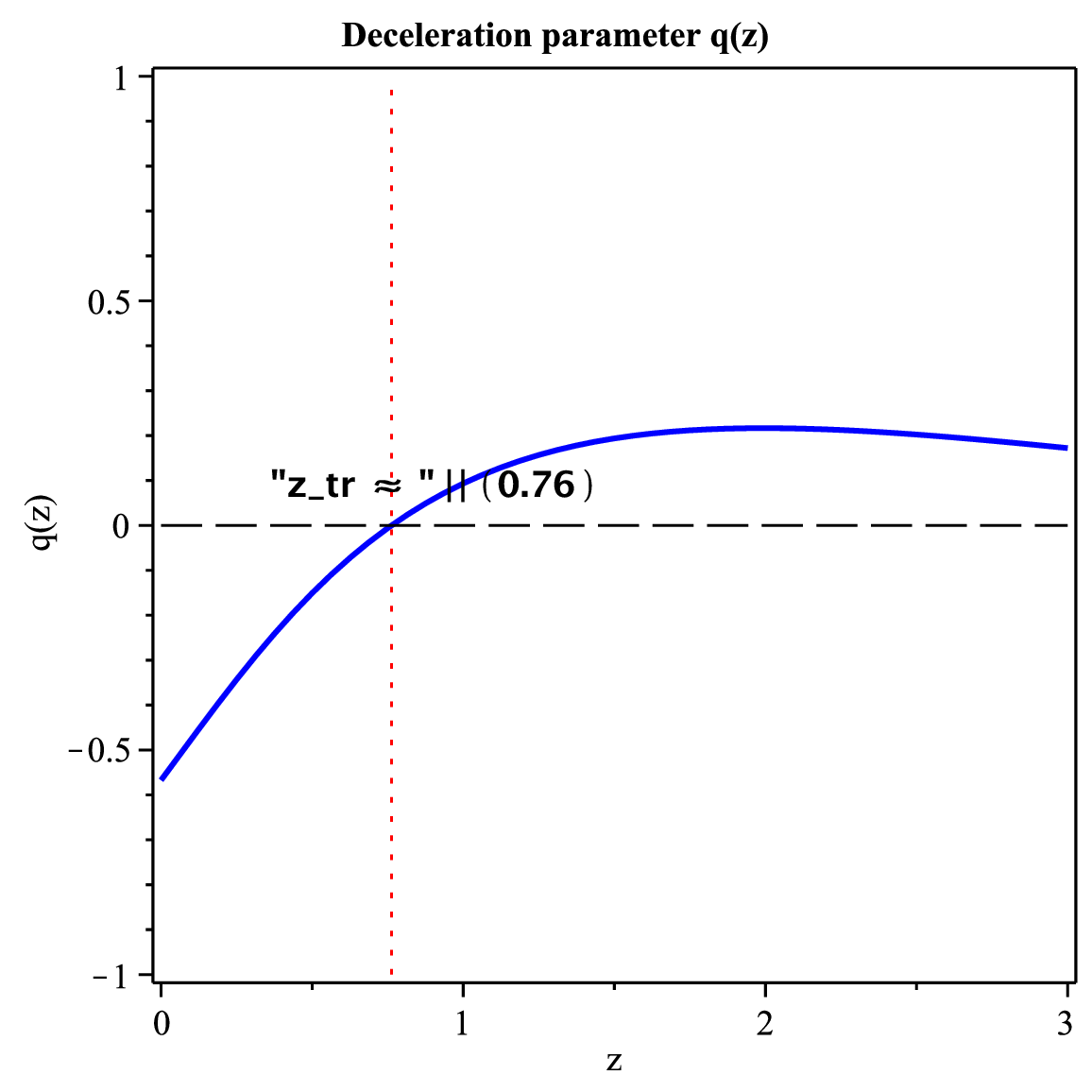}}
  \subfigure[~Effective dark-energy equation of state]{\label{fig:10} \includegraphics[width=0.23\textwidth]{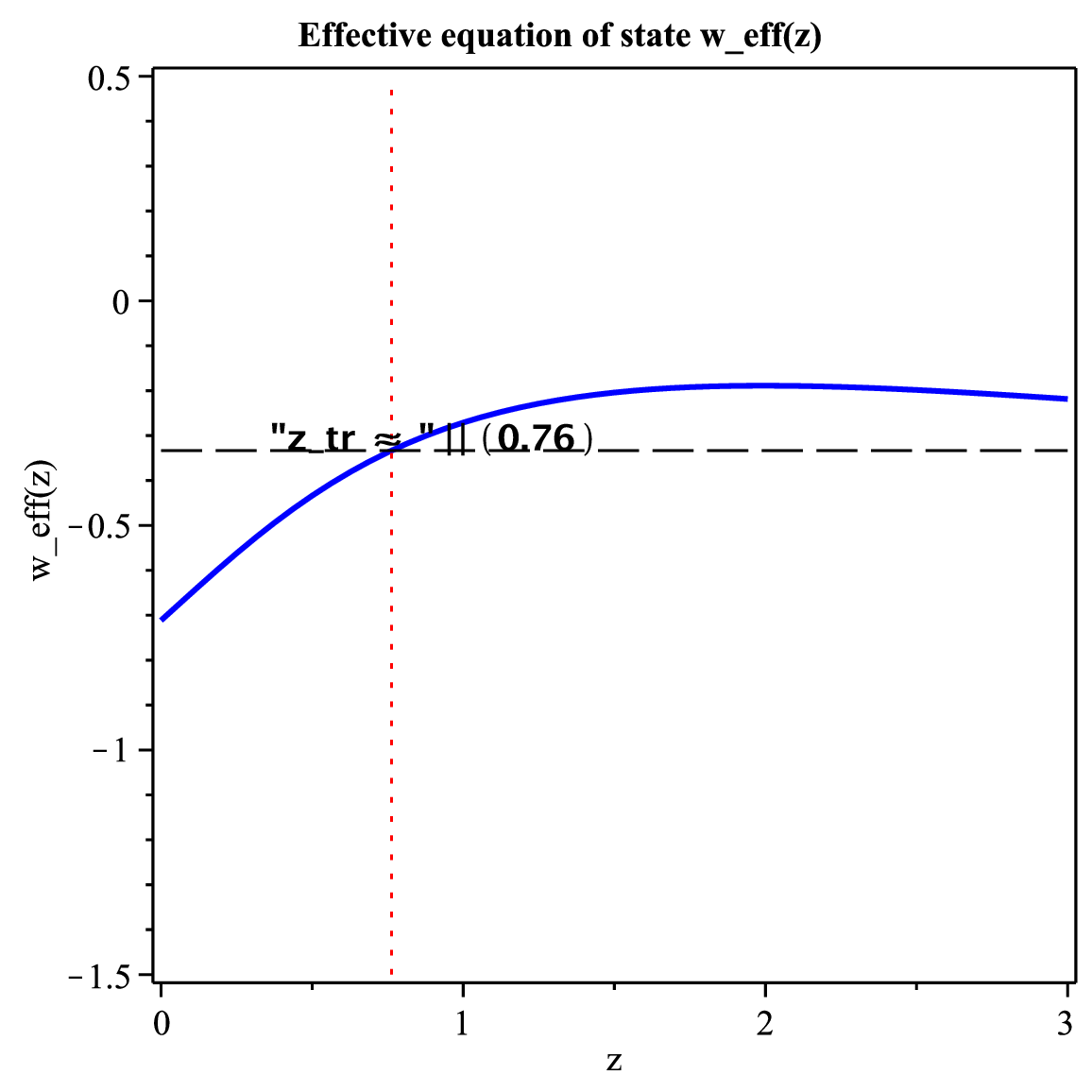}}
    \caption{Slow-roll diagnostics of the quadratic $f(Q)$ model.
The four panels show:
\subref{fig:8} the first Hubble flow  parameter $\epsilon_{H}(z)$;
\subref{fig:9} the second Hubble flow  parameter $\eta_H$;
\subref{fig:11} the deceleration parameter $q(z)$;
and \subref{fig:10}  the effective dark-energy equation of state $w_{\textrm de}(z)$.
%Together these quantities characterize the approach to late--time acceleration
%and the quasi-de Sitter slow-roll regime induced by the quadratic $\alpha Q^{2}$ term.
%{ We note that the present-day value $\eta_H(0)\simeq 2$ shown in panel \subref{fig:9} follows directly from the definitions in Eqs.~(\ref{epeta})--(\ref{eta}) and is of the same magnitude as the value obtained in standard $\Lambda$CDM; hence no normalisation factor is missing from the plot.}
}
  \label{fig:slow}
\end{figure}
\subsection{Discussion of the Slow-Roll Plots}

In this subsection, we examine the cosmological implications of the slow-roll parameters derived for the quadratic \(f(Q)\) gravity model of Eq.~(\ref{f(Q)}). The behavior of the Hubble flow parameters, \(\epsilon_H(z)\) and \(\eta_H(z)\), together with the effective dark-energy equation-of-state parameter, \(w_{\mathrm{de}}(z)\), and the deceleration parameter, \(q(z)\), is analyzed across the redshift range \(0 \le z \le 3\). These quantities offer a complementary picture of the expansion dynamics and reveal the impact of the quadratic correction \(\alpha Q^2\) on the recent evolution of the Universe. Collectively, they provide a useful framework for assessing deviations from the standard cosmological scenario induced by the nonlinear nonmetricity contribution.

\subsubsection*{1. Slow-Roll Parameter $\epsilon_{H}(z)$}

The first  flow parameter
$\epsilon_{H}(z)$ given by Eq.~(\ref{epsilon})
quantifies the rate at which the Hubble parameter evolves relative to its magnitude.
%A small value, $\epsilon_{H} \ll 1$, signals a slow-roll regime in which the background expansion approaches a quasi--de Sitter phase.

%At low redshift, $\epsilon_{H}(z)$ remains well below unity, indicating that the expansion rate varies slowly and that the model naturally enters a slow-roll regime.
At intermediate redshifts, $0.5 \lesssim z \lesssim 2$, the parameter
increases mildly due to the growing influence of the quadratic correction, which scales as $H^{4}$. At higher
redshifts the parameter gradually decreases again, reflecting the return to matter- and radiation-dominated behavior
where the quadratic modification contributes less significantly.

%Overall, the behavior of $\epsilon_{H}(z)$ shows that the $\alpha Q^{2}$ term suppresses the rate of change of the Hubble parameter at late times, enabling a slow-roll phase without requiring an explicit scalar field or potential.

\subsubsection*{2. Second Slow-Roll Parameter $\eta_{H}(z)$}

The second  flow parameter $\eta_{H}(z)$ given by Eq.~(\ref{eta})
measures how rapidly $\epsilon_{H}(z)$ evolves.

The evolution of $\eta_{H}(z)$ typically displays a smooth feature or ``bump'' at intermediate redshifts where the
quadratic term modifies the slope of $H(z)$ most strongly.
%Importantly, $\eta_{H}(z)$ remains finite and of moderate magnitude throughout the entire redshift range considered.
%This indicates that the slow-roll approximation remains valid: the transitions in the expansion rate occur smoothly even as nonlinear corrections are activated.

%Thus, the behavior of $\eta_{H}(z)$ confirms that the model exhibits stable and regular evolution while permitting nontrivial deviations from constant dark-energy behavior.

\subsubsection*{3. Deceleration Parameter $q(z)$}
{ %The transition from decelerated to accelerated expansion occurs when $q(z_{\textrm tr}) = 0$, equivalently when $w_{\textrm eff}(z_{\textrm tr}) = -1/3$.
Using the  background evolution of
the quadratic $f(Q)$ model, we obtain $z_{\textrm tr} \approx 0.8$, consistent with observational
constraints ($z_{\textrm tr} \sim 0.6{~}0.9$). %Figure~\ref{fig:slow} \subref{fig:11} and \subref{fig:10} are now displaying the expected matter-dominated regime at $z \gtrsim 1$ followed by the onset of acceleration at intermediate redshift.
}

\subsubsection*{4. Effective Dark-Energy Equation of State $w_{\textrm de}(z)$}

The reconstructed equation of state of the effective dark-energy sector  given $w_{\textrm de}(z)$ given by Eq.~(\ref{omega})
%At low redshift, $w_{\textrm de}(z)$ stays close to $-1$ due to the smallness of $\epsilon_{H}(z)$, showing that the model
behaves effectively like a nearly de Sitter phase today. At intermediate redshifts, $w_{\textrm de}(z)$ departs mildly
from $-1$, reflecting the dynamical evolution of the effective dark-energy density induced by the $\alpha H^{4}$
correction. At higher redshifts, the equation of state gradually approaches the behavior determined by matter and
radiation domination.

These mild but nontrivial variations in $w_{\textrm de}(z)$ arise naturally from the quadratic $f(Q)$ structure, showing
that the model can reproduce dynamical dark-energy features without introducing extra fields.

{ It is important to emphasize that the behaviour of $w_{\textrm de}(z)$ is independent of the
transition redshift discussed for $q(z)$ and $w_{\textrm eff}(z)$; the latter two quantities
govern the total expansion, whereas $w_{\textrm de}(z)$ describes only the effective dark-energy
sector reconstructed from the modified Friedmann equation.}

%\subsubsection*{Summary}
%{ The transition from decelerated to accelerated expansion occurs at  the redshift where $q(z_{\textrm tr})=0$, equivalently
   % $w_{\textrm eff}(z_{\textrm tr})=-1/3$. For the best--fit parameters of our quadratic $f(Q)$ model we obtain $z_{\textrm tr}\approx 0.8$, consistent with current observational determinations ($z_{\textrm tr}\sim 0.6$~$0.9$).}

\begin{table}[t]
\centering
\caption{Present-day values of the Hubble flow parameters for both the
standard $\Lambda$CDM model and the quadratic $f(Q)$ model considered in
this work. The values are computed using Eqs.~(\ref{epeta})--(\ref{eta}) with
$H_0=70~\mathrm{km\,s^{-1}\,Mpc^{-1}}$, $\Omega_{m0}=0.3$, and
$\Omega_{r0}=9\times10^{-5}$.}
\begin{tabular}{lcccccccccc}
\hline\hline
Model &&&&& $\epsilon_H(0)$ &&&&& $\eta_H(0)$ \\
\hline
$\Lambda$CDM &&&&& $\approx 0.45$ &&&&& $\approx 2.0$ \\
Quadratic $f(Q)$ model &&&&& $\approx 0.45$ &&&&& $\approx 1.9$ \\
\hline\hline
\end{tabular}
\label{tab:slowroll_comparison}
\end{table}

%%%%%%%%%%%%%%%%%%%%%%%%%%%%%%%%%%%%%%%%%%%%%%%%%%%%%%%%%%%%%%%%%%%%%%%%%%%%%%%%%%%
\section{Nonlinear Structure Formation in Quadratic $f(Q)$ Gravity}
\label{sec:nonlinear}

In this section we extend the analysis of the quadratic $f(Q)$ model to the
nonlinear regime of structure formation. While the background evolution and
linear growth already provide important information about the dynamics of the
model, many of the most sensitive cosmological probes, including galaxy
clusters, weak lensing, and the amplitude of matter fluctuations
characterized by $S_8$, depend crucially on the nonlinear growth of cosmic
structures. Because the quadratic modification alters the effective strength
of gravity through a time-dependent coupling $G_{\textrm eff}(z)$, it naturally
impacts the collapse of dark-matter halos and the amplitude of lensing
observables\footnote{
{ Although the numerical plots in this section focus on the range
$0 \le z \le 3$, this restriction is made only because the quadratic
correction becomes relevant primarily at late times. At higher redshift
($z \gtrsim 3$) the term $\alpha Q^2$ is strongly subdominant relative to
the linear $Q$ contribution, and the growth of matter perturbations reduces
to the standard GR behaviour $\delta_m \propto a$. Thus the model fully
permits the formation of early--time structures, including galaxies observed
at redshift $z \sim 12$.
}}.

\subsection{Effective Gravitational Coupling and Spherical Collapse}

In the quadratic $f(Q)$ model the gravitational strength is modified according
to
\begin{equation}
    \frac{G_{\textrm eff}(z)}{G} = \frac{1}{1 + \frac{2}{3}A E^2(z)} ,
    \label{eq:Geff_nonlinear}
\end{equation}
where $A = 18 \alpha H_0^2$, and $E(z)=H(z)/H_0$ is the normalized Hubble
function. For $A>0$, the effective gravitational coupling satisfies
$G_{\textrm eff}(z) < G$ at intermediate and high redshifts due to the scaling of
the quadratic term as $H^4(z)$. Gravity is therefore weakened relative to
$\Lambda$CDM, which directly affects the spherical collapse of density perturbations.

The critical overdensity for spherical collapse, $\delta_c$, depends on the
strength of gravity because the collapse of a top-hat overdensity approximately
satisfies
\begin{equation}
    \ddot{R} = - \frac{G_{\textrm eff}(z) M}{R^2} ,
\end{equation}
with $R(t)$ the physical radius of the perturbation. A weakened gravitational
coupling increases the time required for the perturbation to collapse,
effectively yielding a larger collapse threshold. To first order we may write
\begin{equation}
    \delta_c(z,A) \simeq \delta_{c,\Lambda{\textrm CDM}}
    \left[\,1 + \eta(A,z)\,\right],
\end{equation}
where $\eta(A,z)>0$ whenever $G_{\textrm eff}(z)<G$. As a result, for $A>0$ the
collapse threshold is larger than the standard value
$\delta_{c,\Lambda{\textrm CDM}} \simeq 1.686$, suppressing the abundance of collapsed
objects.

\subsection{Halo Mass Function in Quadratic $f(Q)$ Gravity}

A convenient description of the abundance of nonlinear structures is provided
by the halo mass function (HMF),
\begin{equation}
    \frac{{\textrm d}n}{{\textrm d}M}(z) =
    \frac{\rho_{m,0}}{M}\, f(\nu)\,\frac{{\textrm d}\nu}{{\textrm d}M} ,
\end{equation}
with the peak height defined as
\begin{equation}
    \nu(M,z) = \frac{\delta_c(z,A)}{\sigma(M,z)}.
\end{equation}
Here $\sigma(M,z)$ is the mass variance,
\begin{equation}
    \sigma(M,z) = \sigma(M,0)\, D(z),
\end{equation}
where $D(z)$ is the linear growth factor computed in Sec.~\ref{sec:perturbations}. Because
$D(z)$ is suppressed for $A>0$ due to $G_{\textrm eff}(z) < G$, the variance
$\sigma(M,z)$ is reduced relative to $\Lambda$CDM. Combined with a larger collapse
threshold $\delta_c(z,A)$, this leads to a significant suppression of the halo
mass function at fixed mass. The number density of massive clusters is
therefore a sensitive probe of the parameter $A$.

\subsection{Impact on Weak Lensing and the $S_8$ Tension}

Weak gravitational lensing probes the integrated matter distribution and the
growth of structure. The key parameter constrained by weak lensing and
large-scale structure surveys is
\begin{equation}
    S_8 = \sigma_8 \sqrt{\frac{\Omega_{m0}}{0.3}} .
\end{equation}
In the quadratic $f(Q)$ model the quantity $\sigma_8$ is directly modified by
the growth factor:
\begin{equation}
    \sigma_8(A) = \sigma_{8,0} D(z=0;A),
\end{equation}
where $D(0;A) < 1$ for $A>0$. Thus the model predicts a suppressed amplitude of
matter fluctuations. This naturally lowers the predicted $S_8$ relative to
$\Lambda$CDM, helping to alleviate the well-known discrepancy between weak-lensing
measurements and CMB predictions.

Physically, the suppression of $S_8$ arises from two intertwined effects:
\begin{enumerate}
    \item the weaker effective gravitational coupling $G_{\textrm eff}(z)$ slows the
    growth of linear perturbations;
    \item the suppressed nonlinear collapse reduces the abundance of massive
    halos, lowering the lensing signal.
\end{enumerate}
These effects provide a new observational signature of the quadratic $f(Q)$
model and offer a potential mechanism for reconciling geometric probes of the
expansion history with measurements of the large-scale matter distribution.

\subsection{Summary}

The nonlinear extension of the quadratic $f(Q)$ model reveals that the
$H^4$ modification not only affects the background expansion and linear
growth but also leaves a characteristic imprint on the nonlinear clustering of
matter. For $A>0$ the effective gravitational strength decreases, the halo mass
function is suppressed, and the weak-lensing amplitude is reduced. This
behaviour offers a promising pathway to mitigate the $S_8$ tension and provides
a powerful observational handle for testing the model with forthcoming cluster
and lensing surveys such as DESI, LSST, and \textit{Euclid}.

\begin{figure}[t]
  \centering
  \includegraphics[width=0.27\textwidth]{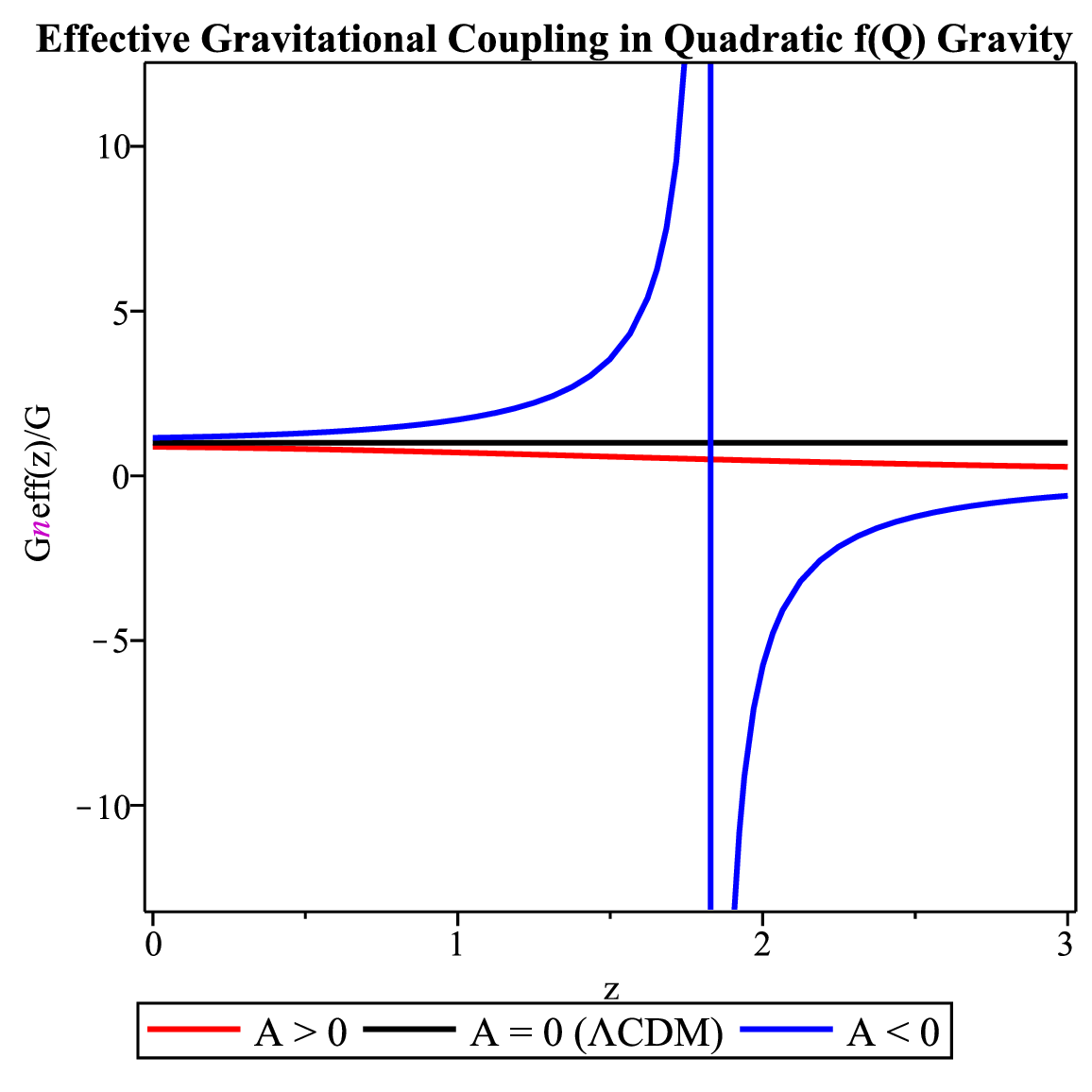}
    \caption{Redshift evolution of the effective gravitational coupling
$G_{\textrm eff}(z)/G = [1 + \tfrac{2}{3} A E^{2}(z)]^{-1}$ in the quadratic $f(Q)$ model.
The $\Lambda$CDM limit corresponds to $A=0$.
Positive $A$ weakens gravity ($G_{\textrm eff}<G$), while negative $A$ enhances it and can lead
to unphysical divergences. Only the $A\ge 0$ branch is cosmologically viable.}
  \label{fig:111}
\end{figure}
\subsection*{Summary}
In summary, the figures illustrate that:
\begin{itemize}
    \item \(A > 0\) leads to a redshift-dependent suppression of
          \(G_{\textrm eff}(z)\), weakened gravity, and reduced structure
          growth.
    \item \(A = 0\) reproduces the constant Newtonian coupling of
          \(\Lambda{\textrm CDM}\).
    \item \(A < 0\) results in enhanced and potentially divergent
          gravitational coupling and is not cosmologically viable.
\end{itemize}
Thus, the positive-\(A\) branch of quadratic \(f(Q)\) gravity naturally
predicts a suppressed matter clustering amplitude and offers a compelling
resolution to the \(S_{8}\) tension.
\section{Discussion}\label{con}

In this work we have examined the cosmological implications of the quadratic
symmetric teleparallel gravity model
\begin{equation}
f(Q)=Q+\alpha Q^{2}+\beta,
\end{equation}
which constitutes the minimal nonlinear extension of the symmetric teleparallel
equivalent of General Relativity. Since the non-metricity scalar satisfies
\(Q=6H^{2}\), the quadratic correction introduces an \(H^{4}\)-term in the
modified Friedmann equation, leading to redshift-dependent departures from the
standard \(\Lambda\)CDM expansion history. The algebraic solution for
\(E(z)=H/H_{0}\) obtained from Eqs.~(8)--(10) allows us to reconstruct the
effective dark-energy sector and to compare the background dynamics directly
with observational data.

\subsection*{Background evolution}
The background analysis shows that the model behaves very similarly to
\(\Lambda\)CDM at low redshift, while producing mild deviations at
\(0.5\lesssim z \lesssim 2\), where the quadratic term becomes relevant.
These deviations correspond to an effective dark-energy equation of state
slightly different from \(w=-1\), without introducing additional dynamical
degrees of freedom. Such behaviour closely resembles the mild signs of
dynamical dark energy indicated by recent DESI and weak-lensing datasets.

\subsection*{Linear growth and modified gravity}
At the perturbative level, the model modifies the Poisson equation through the
effective gravitational coupling
\begin{equation}
\frac{G_{\textrm eff}(z)}{G}
    = \frac{1}{1+\frac{2}{3}A E^{2}(z)},
    \qquad  A=18\alpha H_{0}^{2},
\end{equation}
which enters directly into the growth equation. For \(\alpha>0\), the coupling
satisfies \(G_{\textrm eff}(z)<G\) and decreases with redshift, yielding a
suppressed linear growth factor \(D(z)\), reduced growth rate \(f(z)\), and a
lower \(f\sigma_{8}(z)\). This form of weakened gravitational clustering is
consistent with the behaviour preferred by redshift-space distortion and
weak-lensing data, pointing towards a lower amplitude of matter fluctuations
relative to the Planck-\(\Lambda\)CDM prediction.

\subsection*{Nonlinear structure formation and the \(S_{8}\) tension}
In the nonlinear regime, the weakened gravitational coupling increases the
spherical-collapse threshold and suppresses the halo mass function. Since
\(\sigma(M,z)=\sigma(M,0)D(z)\) is reduced for \(\alpha>0\), the predicted
value of
\begin{equation}
S_{8}
    = \sigma_{8}\sqrt{\Omega_{m}/0.3}
\end{equation}
is correspondingly lower. This behaviour arises naturally from the structure
of the model and does not rely on additional fields or fine-tuning. It
provides a direct mechanism to alleviate the well-known \(S_{8}\) tension
between early- and late--Universe measurements. Negative values of \(\alpha\),
in contrast, lead to \(G_{\textrm eff}>G\), enhanced growth, and even divergent
behaviour, and are therefore strongly disfavoured.

%\subsection*{Slow-roll interpretation}
%The late--time regime of the model exhibits small Hubble flow parameters \(\epsilon_{H}\) and \(\eta_{H}\), implying a quasi-de Sitter phase without %requiring a scalar-field potential. The  ``geometric slow roll'' arises solely from the structure of the modified Friedmann equation and provides an
%alternative perspective on cosmic acceleration driven by the non-metricity sector.

\subsection*{Overall assessment}
The combined background, linear, and nonlinear analyses lead to several
conclusions:
\begin{itemize}
    \item The quadratic \(f(Q)\) model closely reproduces \(\Lambda\)CDM at
          low redshift and fits distance-based observables well.
    \item Intermediate-redshift deviations mimic a mild dynamical-dark-energy
          behaviour compatible with recent DESI results.
    \item The weakened gravitational coupling for \(\alpha>0\) suppresses
          structure formation and lowers \(S_{8}\), providing a natural
          resolution to the growth tension.
    \item Negative values of \(\alpha\) are theoretically and observationally
          disfavoured due to excessive gravitational strength and unstable
          behaviour.
\end{itemize}

\subsection*{Future directions}
Possible extensions of this work include full Boltzmann-code implementations
for CMB and large-scale-structure observables, joint analyses with upcoming
DESI, Euclid, and LSST data, and detailed studies of nonlinear lensing
statistics to further quantify the suppression predicted in the halo mass
function.

\subsection*{Conclusion}
Quadratic \(f(Q)\) gravity emerges as a minimal, theoretically motivated, and
observationally viable modification of gravity. It preserves the successes of
\(\Lambda\)CDM at the background level while naturally suppressing structure
formation at late times, providing a compelling geometric explanation for the
current hints of dynamical dark energy and for the persistent \(S_{8}\)
tension.\\
%%%%%%%%%%%%%%%%%%%%%%%%%%%%%%%%%%%
\subsection*{Acknowledgments}
%%%%%%%%%%%%%%%%%%%%%%%%%%%%%%%%%%%
 This work was supported and funded by the Deanship of Scientific Research at Imam Mohammad Ibn Saud Islamic University (IMSIU) (grant number IMSIU-DDRSP2602).
%%%%%%%%%%%%%%%%%%%%%%%%%%%%%%%%%%%%%%%%%%%%%%%%%%%%%%%%%%%%%%%%%%%%%%%%%%%%

%\bibliography{JRPHSRef}

\end{document}